\RequirePackage{lineno}
\documentclass[twocolumn,preprintnumbers,amsmath,amssymb,aps]{revtex4}
\usepackage{txfonts}
\usepackage{dsfont}
\usepackage{amssymb}
\usepackage{booktabs}
\usepackage{graphicx}
\usepackage{color}
\usepackage{dcolumn}
\usepackage{amsmath}
\usepackage{epstopdf}
\usepackage{bm}
\usepackage{ulem}
\hfuzz=\maxdimen
\begin{document}
\title{Quantum dynamical characterization and simulation of topological phases with high-order band inversion surfaces}
\author{Xiang-Long Yu$^1$}
\thanks{These authors contributed equally to this work.}
\author{Wentao Ji$^{2,3}$}
\thanks{These authors contributed equally to this work.}
\author{Lin Zhang$^{4,5}$}
\thanks{These authors contributed equally to this work.}
\author{Ya Wang$^{2,3}$}
\email[Corresponding author. E-mail: ]{ywustc@ustc.edu.cn}
\author{Jiansheng Wu$^{1,6}$}
\email[Corresponding author. E-mail: ]{wujs@sustech.edu.cn}
\author{Xiong-Jun Liu$^{4,5,1}$}
\email[Corresponding author. E-mail: ]{xiongjunliu@pku.edu.cn}
\affiliation{1 Department of Physics and Shenzhen Institute for Quantum Science and Engineering, Southern University of Science and Technology, Shenzhen 518055, P. R. China}
\affiliation{2 Hefei National Laboratory for Physical Sciences at the Microscale and Department of Modern Physics, University of Science and Technology of China, Hefei 230026, P. R. China}
\affiliation{3 CAS Key Laboratory of Microscale Magnetic Resonance, University of Science and Technology of China, Hefei 230026, P. R. China}
\affiliation{4 International Center for Quantum Materials, School of Physics, Peking University, Beijing 100871, P. R. China}
\affiliation{5 Collaborative Innovation Center of Quantum Matter, Beijing 100871, P. R. China}
\affiliation{6 Guangdong Provincial Key Laboratory of Quantum Science and Engineering, Shenzhen Institute for Quantum Science and Engineering, Southern University of Science and Technology, Shenzhen 518055, Guangdong, P. R. China}

\begin{abstract}
How to characterize topological quantum phases is a fundamental issue in the broad field of topological matter. From a dimension reduction approach, we propose the concept of high-order band inversion surfaces (BISs) which enable the optimal schemes to characterize equilibrium topological phases by far-from-equilibrium quantum dynamics, and further report the experimental simulation. We show that characterization of a $d$-dimensional ($d$D) topological phase can be reduced to lower-dimensional topological invariants in the high-order BISs, of which the $n$th-order BIS is a $(d-n)$D interface in momentum space. In quenching the system from trivial phase to topological regime, we unveil a high-order dynamical bulk-surface correspondence that the quantum dynamics exhibits nontrivial topological pattern in arbitrary $n$th-order BISs, which universally corresponds to and so characterizes the equilibrium topological phase of the post-quench Hamiltonian. This high-order dynamical bulk-surface correspondence provides new and optimal dynamical schemes with fundamental advantages to simulate and detect topological states, in which through the highest-order BISs that are of zero dimension, the detection of topological phase relies on only minimal measurements. We experimentally build up a quantum simulator with spin qubits to investigate a $3$D chiral topological insulator through emulating each momentum one by one and measure the high-order dynamical bulk-surface correspondence, with the advantages of topological characterization via highest-order BISs being demonstrated.
\end{abstract}


\maketitle

\section{INTRODUCTION}
The bulk-boundary correspondence (BBC) is a fundamental mechanism in topological quantum phases, such as in quantum Hall effect, topological insulators, and topological superconductors, in which the topological number of the bulk links to the number of the robust gapless states on the boundary~\cite{PRL45.494,PRL48.1559,bookTQHE,RMP82.3045,RMP83.1057,Yan2012,Chiu2016,Yan2017,NPB922.62,CMP345.675,JGP124.421,Nat547.298,PU44.131,PRB61.10267,RPP75.76501,RMP87.137,RPP80.076501}.
In consequence, the BBC is widely utilized to identify the topological quantum states and measure the topological invariants, e.g. by transport and angle-resolved photoelectron spectroscopy (ARPES) experiments~\cite{Science318.766,Nat452.970,NP5.598,Science340.167,Science349.613,PRX5.031013,LiuX2019}. Recently, a new development has been made for topological systems, called the higher-order topological phases~\cite{Science357.61,SA4.eaat0346,PRL119.246401,PRL119.246402,PRB92.085126,PRB97.205136,PRX9.011012}.
In such quantum systems, a $d$-dimensional ($d$D) bulk has nontrivial topology in the presence of a certain spatial symmetry, while the corresponding gapless modes do not exist on the ($d-1$)D boundary but survive on a lower ($d-n$)D boundary with $n>1$, due to spatial symmetry breaking on the former but preserving on the later. This gives a high-order BBC in the real space.

While the BBC is well defined in the real space, it seems not straightforward to extend such concept to the momentum space, since the latter is intrinsically closed and has no well-defined boundaries.
However, a new method to classify topological states was proposed recently, showing that the topological number of a $d$D system corresponds to the nontrivial (pseudo)spin texture on the ($d-1$)D momentum subspace, called band inversion surfaces (BISs)~\cite{SB63.1385,PRA99.053606,Zhang2019b,Zhang2020}. Rather than being the boundary, the BISs are interfaces in the momentum space, across which the bulk bands are inverted as the interband couplings are absent. This method is particularly useful in two aspects. First, it enables to characterize topological phases by quantum dynamics, which is a topic attracting fast-growing attention recently~\cite{dynamic1,dynamic2,dynamic3,dynamic4,dynamic5,PRB98.205417,dynamic6,McGinley2019,PRL124.160402}. By suddenly tuning a system from initially trivial phase to topological regime, the quench dynamics exhibits on ($d-1$)D BISs nontrivial dynamical patterns which are related to the $d$D bulk topology, rendering a dynamical bulk-surface correspondence (dBSC) in the momentum space~\cite{SB63.1385,PRA99.053606,Zhang2019b,Zhang2020}. The dBSC opens the way to simulate and detect equilibrium topological phases by far-from-equilibrium quench dynamics, and has been recently actively studied in experiment with ultracold atoms~\cite{PRL121.250403,PRL123.190603,NP15.911,ZWang2020}, solid-state spin systems~\cite{PRA100.052328,Xin2020,Ji2020}, and superconducting curcuits~\cite{Niu2020}. Moreover, as a momentum-space counterpart of the BBC in real space, the dBSC expands the ability of simulating topological phases with the typical quantum simulators like ultracold atoms where the real-space open boundary is hard to construct, but the momentum-space information can be readily measured.
The advantages including the high-precision measurement of the topological phases are also confirmed in experiments based on dBSC compared with traditional methods built on equilibrium theories~\cite{PRL121.250403,PRL123.190603,NP15.911,ZWang2020,PRA100.052328,Xin2020,Ji2020}.

In this work, we propose the concept of high-order BISs for the dynamical characterization and simulation of topological quantum phases by extending the dBSC to high order, and report the experimental observation. The $n$th-order BIS is a $(d-n)$D interface on the $(n-1)$th-order BIS. From a dimension reduction approach, we show that through quantum quenches the bulk topology of a $d$D phase can be uniquely characterized by the dynamical topology emerging on any high-order BISs, manifesting the high-order dBSC. The prediction provides the optimal schemes and considerably expand the freedom for dynamical characterization and simulation of topological phases. Experimentally, we build up a quantum simulator using nitrogen-vacancy (NV) center to investigate the high-order dBSC in a $3$D chiral topological phase through emulating each momentum one by one, with the great advantages of the minimal measurement strategy in the simulation being demonstrated.

The paper is organized as follows. In Sec. II, we introduce the concept of high-order BISs and topological invariants from dimension reduction. In Secs. III and IV, we propose two dynamical schemes for the characterization and simulation of topological phases. In Sec. V, we experimentally build up a quantum simulator using NV center for the dynamical characterization of topological phases. Finally, we conclude in Sev. VI with an outlook. Technical details are provided in Appendices.

\section{High-order band inversion surfaces and topological characterization}

We start with the $d$D topological phases classified by integer invariants in the Altland-Zirnbauer (AZ) symmetry classes, which are winding numbers in odd dimensions and Chern numbers in even dimensions~\cite{PRB78.195125,Chiu2016,Kitaev2009}. The Hamiltonian is described in the form~\cite{PRB88.125129,PRB88.075142},
\begin{eqnarray}\label{eq:general H}
H = \mathbf{h}\left( \mathbf{k} \right) \cdot \mathbf{\gamma}  = \sum\limits_{i = 0}^d {{h_i}\left( \mathbf{k} \right){\gamma _i}},
\end{eqnarray}
where the Clifford algebra matrices $\gamma_{i}$ satisfies anticommunication relation, $\left\{ {{\gamma _i},{\gamma _j}} \right\} = 2{\delta _{ij}}$, and has matrix dimension ${n_d} = {2^{d/2}}$ (or ${2^{(d+1)/2}}$) when the spatial dimension $d$ is even (or odd). For $1$D and $2$D systems, the $\gamma$ matrices are the Pauli ones
and Eq. (\ref{eq:general H}) describes a two-band model, such as the Su-Schrieffer-Heeger chain~\cite{PRB22.2099} and the Haldane model of the integer quantum Hall effect~\cite{PRL61.2015}. For $3$D and $4$D systems, the $\gamma$ matrices are the Dirac ones and the corresponding Hamiltonian describes a four-band model, such as $3$D DIII class superconductors and AIII class topological insulators~\cite{PRB78.195125}, and $4$D quantum Hall insulators~\cite{Science294.823}. In general, the $\gamma$ matrices can be regarded as the spin or pseudospin operators.

\begin{figure}[tbp]
\centering
\setlength{\abovecaptionskip}{2pt}
\setlength{\belowcaptionskip}{4pt}
\includegraphics[angle=0, width=1 \linewidth]{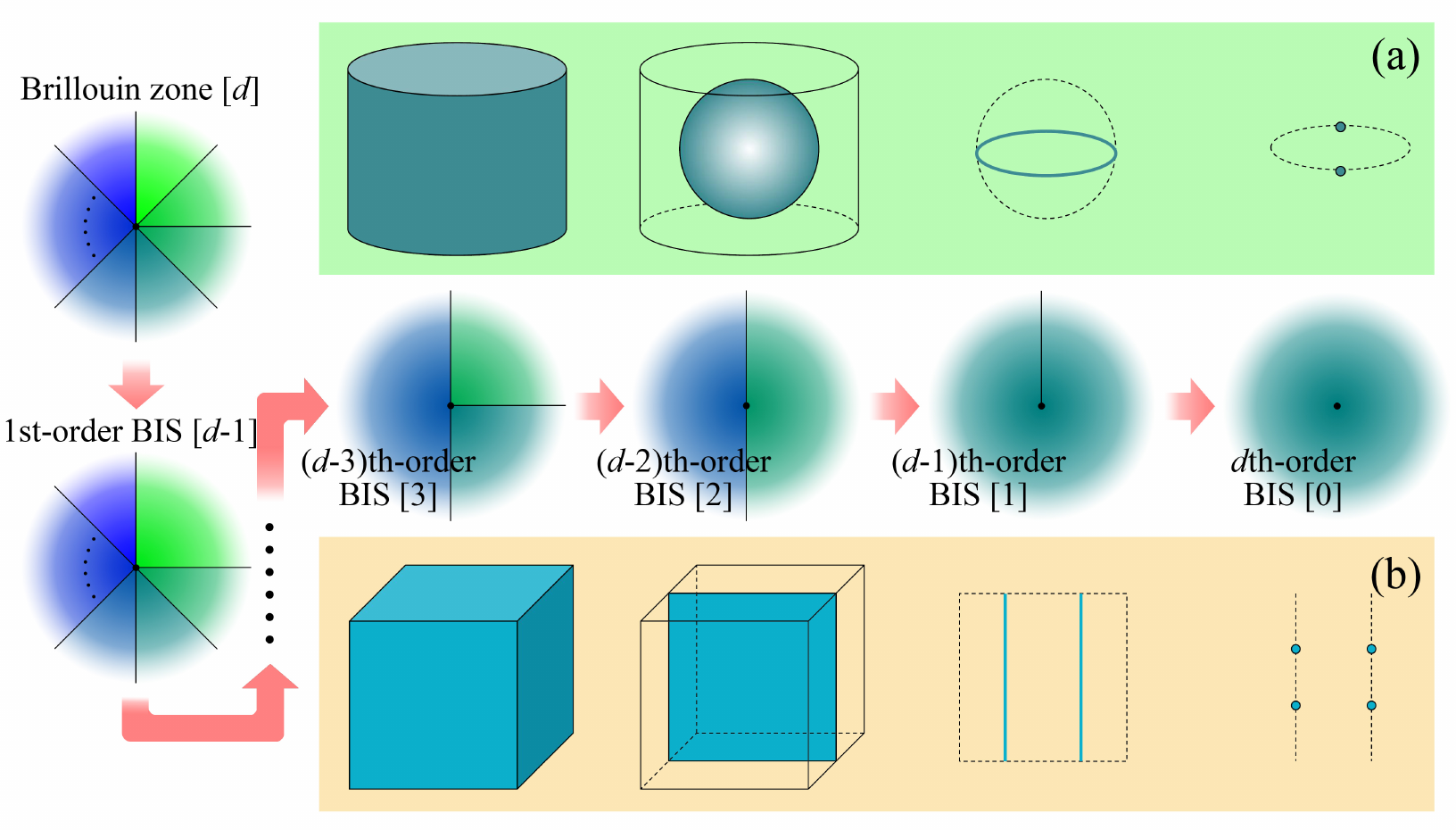}
\caption{A sketch of high-order BISs. A $d$D BZ goes through $d$ dimension reductions and finally becomes a $0$D BIS. Black lines and the number in square brackets correspond to the dimension of the BZ or the BIS. Two insets with (a) green and (b) orange backgrounds illustrate two different kinds of dimension reduction processes of the BIS from three dimensions to zero dimension.
}
  \label{fig:high-order BIS}
\end{figure}

The bulk topology defined in the whole Brillouin zone (BZ) is quantified by a $d$D invariant, and can be reduced to the lower-dimensional topology on the $(d-1)$D BIS which is obtained by choosing one component, e.g. the $i_0$th component of $\mathbf{h}(\mathbf{k})$ to satisfy $h_{i_0}(\mathbf{k})=0$, with $i_0 \in \{0 \sim d\}$, and is a $(d-1)$D closed surface in the BZ~\cite{SB63.1385}. Here we name it as the first-order BIS, $1\text{-BIS}=\{\mathbf{k}\in\mathrm{BZ}\vert h_{i_0}(\mathbf{k})=0\}$, which
can be interpreted in a physically transparent way as follow. We take that the $h_{i_0}$-term characterizes the energy dispersions of the bulk bands, while the remaining terms of the Hamiltonian, denoted by $\mathbf{h}^{(1)}=\{h_{i_1},h_{i_2},\cdots,h_{i_d}\}$, characterize the coupling (pseudospin-orbit coupling) between different bands. Then the 1-BIS is simply a momentum subspace, across which the energy of the half of these bulk bands is inverted with respect to that of the other half before taking into account the interband couplings. The interband coupling term $\mathbf{h}^{(1)}$ opens a gap on the BIS, leading to the gapped topological phase, and moreover, the vector field $\mathbf{h}^{(1)}$ exhibits nontrivial topology on the 1-BIS, which is captured by a $(d-1)$D topological invariant (winding or Chern number) on the 1-BIS. This renders the bulk-surface duality of the free-fermion topological phases with integer invariants~\cite{SB63.1385}.

The central concept we propose here is the high-order BIS which can be introduced based on a dimension reduction method. A key observation is that the interband coupling term $\mathbf{h}^{(1)}$ can be further treated as a $(d-1)$D gapped Hamiltonian with integer invariant defined on the $1\text{-BIS}$. Correspondingly, we can define a higher-order (second-order) BIS on the $1\text{-BIS}$ via another component $h_{i_1}$ of $\mathbf{h}^{(1)}$,
namely $2\text{-BIS}=\{\mathbf{k}\in1\text{-BIS}\vert h_{i_1}(\mathbf{k})=0\}$, which equals to $\{\mathbf{k}\in\mathrm{BZ}\vert h_{i_0}(\mathbf{k})=h_{i_1}(\mathbf{k})=0\}$ and is a $(d-2)$D closed surface. In a similar process we can show that the bulk topology is reduced to the $(d-2)$D invariant of the $(d-1)$D vector field $\mathbf{h}^{(2)}=\{h_{i_2},h_{i_3},\cdots,h_{i_d}\}$ on the $2\text{-BIS}$ (see details in Appendix A). Repeating this dimension reduction we obtain the $n$th-order BIS by
\begin{eqnarray}\label{highorderBIS}
n\text{-BIS}=\{\mathbf{k}\in\mathrm{BZ}\vert h_{i_\alpha}(\mathbf{k})=0; \alpha=0,1,\dots,n-1\},
\end{eqnarray}
where $h_{i_\alpha}$'s are $n$ components of $\mathbf{h}$. Obviously, the $n$-BIS is a $(d-n)$D closed surface and formed by the $\mathrm{k}$ points satisfying $\sum_{\alpha  = 1\sim\left( {n - 1} \right)} {\left| {{h_{{i_\alpha }}}({\bf{k}})} \right|}  = 0$. Across the $n\text{-BIS}$ the band energies switch sign in the absence of $\mathbf{h}^{(n)}=\{h_{i_n},h_{i_{n+1}},\cdots,h_{i_d}\}$.
With the above process the bulk topology then reduces to the $(d-n)$D invariant obtained by $\mathbf{h}^{(n)}$ on the $n\text{-BIS}$ that
\begin{eqnarray}\label{eq:T_BZ=T_nBIS}
  {\mathrm T}_{\mathrm{BZ}}[\mathbf{h}] = {\mathrm T}_{n\text{-BIS}}\bigr[\mathbf{h}^{(n)}\bigr],
\end{eqnarray}
where ${\mathrm T}_{\mathcal{M}}[\bold{f}]=\frac{\Gamma[(m+1)/2]}{2\pi^{(m+1)/2}}\frac{1}{m!}\int_{\mathcal{M}}\hat{
\bf{f}}(\bf{k})[\mathrm{d}\hat{\bf{f}}(\mathbf{k})]^m$ is the topological invariant of the vector field $\bf{f}$ on the $m$D manifold $\mathcal{M}$ (see details of proof in Appendix A) and characterizes the integer times that $\hat{\mathbf{f}}$ covers over the corresponding spherical surface $S^{m}$ when $\mathbf{k}$ runs over $\mathcal{M}$.
Eq. (\ref{eq:T_BZ=T_nBIS}) shows a correspondence that maps the classification of bulk topology to the characterization on the $n\text{-BIS}$. This correspondence is valid for any high-order BISs. In addition,
two points are worthwhile to mention for the dimension reduction process. Firstly, the configurations of the high-order BISs are sharply different if choosing different components ($h_{i}$) of the Hamiltonian for the definition in Eq.~\eqref{highorderBIS}, with two typical cases being illustrated in Fig.~\ref{fig:high-order BIS} (a) and (b), respectively, where we particularly sketch the dimension reduction from $3$D to $0$D in the last three steps. Secondly, whenever the BIS of any certain order does not exist in the BZ, the corresponding system is topologically trivial.

The highest-order $d$-BIS is of zero dimension and is obtained for $n=d$, and consists of a finite number of momentum points. Thus the corresponding topological invariant
${\mathrm T}_{d\text{-BIS}}[h_{i_d}]$ only depends on a finite number of momentum points
\begin{eqnarray}\label{eq:T_0_BIS}
{\mathrm T}_{d\text{-BIS}}[h_{i_d}]
   = \frac{1}{2}\sum\limits_{d\text{-BIS}_j} {\left[ {{\mathop{\rm sgn}} \left( {{h_{i_d,{L_j}}}} \right) - {\mathop{\rm sgn}} \left( {{h_{i_d,{R_j}}}} \right)} \right]},
 \end{eqnarray}
 where $j$ corresponds to different sectors of the $d$-BIS. The subscripts $L_j$ and $R_j$ are the left- and right-hand points of the $j$th sector, respectively. They appear in pairs due to the fermion doubling theorem~\cite{FD}.
The \textit{Left} and the \textit{Right} depend on the direction of the integral path on the $1$D ($d-1$)-BIS.
This formula is similar to the Brouwer degree of the map ${\mathbf h}({\mathbf k})$ from ${\mathbf k}$ space to ${\mathbf{h}}$ space~\cite{PKNAWSA11.788} and it maps the classification of bulk topology to the characterization in zero dimension. In this case,
the computation of the topological invariant can be greatly optimized. In the following we further apply the dimension reduction based on high-order BISs to the dynamical characterization of topological phases.

\section{Dynamical characterization and simulation:~scheme I}
We turn to dynamical characterization of topological phases by showing the high-order dBSC and present quantum simulation with concrete models.
The quantum dynamics are induced by quenching an initially fully polarized trivial phase to topological regime. The pre- and post-quench Hamiltonians are $H_{\rm pre}=H_\mathbf{k}+\delta m_j\gamma_j$ with $\delta m_j\gg \left\| {H_\mathbf{k}} \right\|$ and $H_{\rm post}=H_\mathbf{k}$, respectively, where $j$ denotes the quench axis.
Hence, the initial phase is the trivial ground state of $\delta {m_j}{\gamma _j}$.
The spin precesses with respect to the vector field $\bf h$ of $H_{\rm post}$ after quenching.
The characterization is captured by three main steps: (i) Choose a certain direction to perform a deep quench and measure the spin polarization along the same direction. Then the $1$-BIS is formed by the momenta with vanishing time-averaged spin polarization (TASP), given by
\begin{eqnarray}\label{eq:time-averaged SP}
{\overline {\left\langle {{\gamma _i}} \right\rangle } _j}(\mathbf{k})= \mathop {\lim }\limits_{\mathbb{T} \to \infty } \frac{1}{\mathbb{T}}\int_0^\mathbb{T} {dt{\rm{Tr}}\left[ {{\rho _j}\left( {0} \right){e^{iH_\mathbf{k}t}}{\gamma _i}{e^{ - iH_\mathbf{k}t}}} \right]},
\end{eqnarray}
where $i$ and $j$ correspond to the spin polarization and quench axes, respectively, and ${\rho _j}\left( {0} \right)$ is the density matrix of the initial state at the time $t = 0$. The Eq. (\ref{eq:time-averaged SP}) can be simplified as ${\overline {\left\langle {{\gamma _i}} \right\rangle } _j} =  - {h_i} {h_j}/\sum_{i' = 0}^d {h_{i'}^2}$, and the $1$-BIS is determined by ${\overline {\left\langle {{\gamma _j}} \right\rangle } _j}=0$ which corresponds to $h_j = 0$. On the $1$-BIS the vector field $\mathbf{h}^{(1)}$ is perpendicular to the initial spin polarization, leading to a spin precession within the plane perpendicular to $\mathbf{h}^{(1)}$.
(ii) Repeating the quench process with respect to a new axis $\gamma_{j'}$, one measures another $1$-BIS defined by ${\overline {\left\langle {{\gamma _{j'}}} \right\rangle } _{j'}}=0$. The intersection between the two $1$-BISs gives the $2$-BIS. With this process we can perform sequential dimension reduction of the BISs, and determine the $n$-BIS by
$n\text{-BIS}=\{\mathbf{k}\vert {\overline {\left\langle {{\gamma _0}} \right\rangle } _0}={\overline {\left\langle {{\gamma _1}} \right\rangle } _1}=\cdots={\overline {\left\langle {{\gamma _{n-1}}} \right\rangle } _{n-1}}=0\}$,
where without losing generality we have set the quench directions $i = 0$ to $n-1$ in turn, and in each quench only the $i$th spin component needs to be measured. Note that to determine the $j$-BIS, the spin dynamics are need to be measured only on the subspace of $(j-1)$-BIS, so the number of momentum points need to be measured decreases rapidly.
(iii) One can verify that at the momentum $\mathbf{k}$ away from BISs the TASP is finite. Measure spin polarizations along the remaining $d-n+1$ directions near the $n$-BIS to obtain a dynamical spin texture (DST) on the $n$-BIS. Its components are given by
\begin{eqnarray}\label{eq:g}
g_{l}^{n\rm th}\left( \mathbf{k} \right) =  \frac{1}{{{\Upsilon _\mathbf{k}}}}\frac{{\partial {{\overline {\left\langle {{\gamma _l}} \right\rangle}_{n-1} }}}}{{\partial {k_ \bot }}}, \ l=n, n+1, \cdots, d.
\end{eqnarray}
Here ${\Upsilon _k}$ is a normalization factor and $k_\bot$ denotes the momentum perpendicular to the $n$-BIS within the $(n-1)$-BIS and pointing to $h_{n-1}>0$. Physically, the DST describes the variation of the remaining TASP components ${\overline {\langle {\gamma _l}\rangle}}_{n-1}$ across the $n$-BIS. On the $n$-BIS we further find that the DST $\mathbf{g}^{n\rm th}(\mathbf{k})=\hat{\mathbf{h}}^{(n)}(\mathbf{k})$, which is a key result implying that the topology on $n$-BIS is captured by the DST. Then the topological invariant is intuitively described by the coverage of the DST $\mathbf{g}^{n\rm th}(\mathbf{k})$ over the corresponding $(d -n)$D spherical surface $S^{d-n}$ formed by the unit vector $\hat{\mathbf{h}}^{(n)}(\mathbf{k})$ when $\bold k$ runs over the $n$-BIS.
In particular, for $n=d$ we obtain
${\mathrm T}_{d\text{-BIS}}[\mathbf{g}]
   = \frac{1}{2}\sum\limits_{d\text{-BIS}_j} {\left[ {{\mathop{\rm sgn}} \left( {{g^{d\rm th}_{d,{L_j}}}} \right) - {\mathop{\rm sgn}} \left( {{g^{d\rm th}_{d,{R_j}}}} \right)} \right]}$.

\begin{figure}[tbp]
\centering
\setlength{\abovecaptionskip}{2pt}
\setlength{\belowcaptionskip}{4pt}
\includegraphics[angle=0, width=1 \columnwidth]{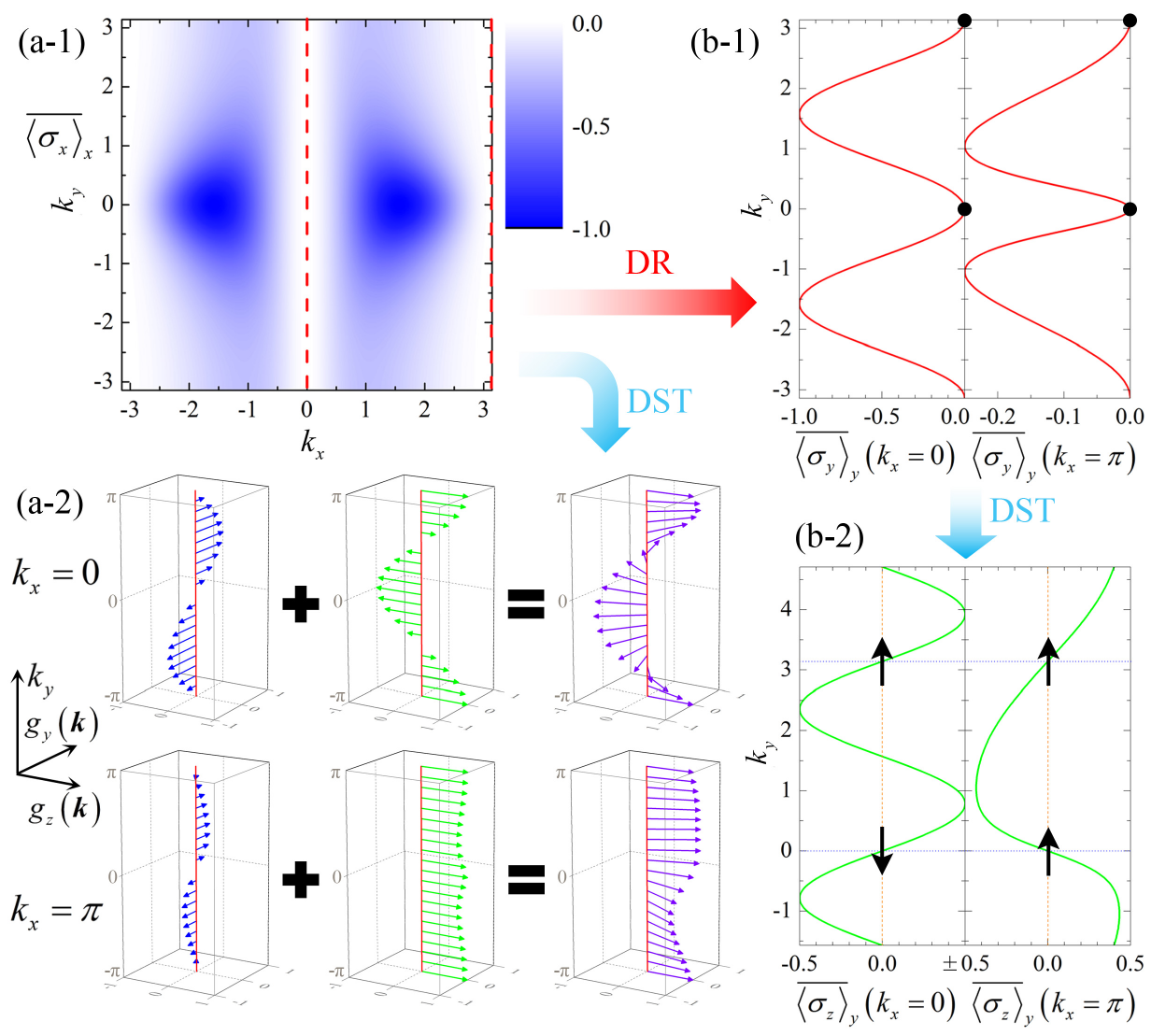}
\caption{(a-1) TASP ${\overline {\left\langle {{\sigma _x}} \right\rangle } _x}$ in the BZ after a quench from a large $m_x$ to $0$. The $1$-BIS is marked by red dashed lines (${k_x}=0$ and $\pi$).
(a-2) DSTs $g^{1st}_y$, $g^{1st}_z$, and $\mathbf{g}^{1st}$ along the $k_y$ direction on the $1$-BIS.
(b-1) ${\overline {\left\langle {{\sigma _y}} \right\rangle } _y}$ on the $1$-BIS after a quench from a large $m_y$ to $0$. The $2$-BIS is marked by four black points $\left( {{k_x},{k_y}} \right) = \{\left( {0,0} \right),\left( {0,\pi } \right),\left( {\pi ,0} \right),\left( {\pi ,\pi } \right)\}$.
(b-2) ${\overline {\left\langle {{\sigma _z}} \right\rangle } _y}$ on the $1$-BIS after a quench from a large $m_y$ to $0$. Black arrows represent the DSTs $\mathbf{g}^{2nd}$ on the $2$-BIS. Here (and in the following figures) red and blue arrows correspond to the dimension reduction (DR) of BIS and the calculation of DSTs, respectively.}
  \label{fig:2D_BIS_a}
\end{figure}

We present the quantum simulation with two realistic models to show explicitly how the high-order dBSC can provide dynamical characterization of topological phases based on dimension reduction. First we consider a $2$D quantum anomalous Hall system with Hamiltonian
\begin{equation}\label{eq:2D H}
H_{2D} = {h_x}{\sigma _x} + {h_y}{\sigma _y} + {h_z}{\sigma _z}
\end{equation}
with ${h_x} = {m_x} + t_{so}^x\sin {k_x}$, ${h_y} = {m_y} + t_{so}^y\sin {k_y}$, ${h_z} = {m_z} - {t_0}\cos {k_x} - {t_0}\cos {k_y}$, and $t_{so}^{x,y} = {m_z} = {t_0}$. This model has been widely realized in experiments~\cite{Wu2016,Yi2019,PRL121.250403,PRL123.190603}.

We first deeply quench the $h_x$ axis from a large $m_x$ to $0$ with $m_{y} = 0$, and measure TASP ${\overline {\left\langle {{\sigma _x}} \right\rangle } _x}$ in the whole BZ. Then the $1$-BIS with ${\overline {\left\langle {{\sigma _x}} \right\rangle } _x} = 0$ can be obtained, which are two lines, ${k_x} = 0$ and $\pi$, as shown in Fig. \ref{fig:2D_BIS_a} (a-1).
On the $1$-BIS we further set $m_{x} = 0$, deeply quench the $h_y$ axis, and measure ${\overline {\left\langle {{\sigma _y}} \right\rangle } _y}$. Then the $2$-BIS which is formed by four momentum points can be obtained by $2\text{-BIS}=\{\mathbf{k}\vert \overline {\left\langle {{\sigma _x}} \right\rangle } _x = 0, \overline {\left\langle {{\sigma _y}} \right\rangle } _y=0\}$,
and is marked in Fig. \ref{fig:2D_BIS_a} (b-1).
With the same quench we could measure the TASP ${\overline {\left\langle {{\sigma _z}} \right\rangle } _y}$ around the four $k$ points and calculate the corresponding DST [defined in Eq. (\ref{eq:g})] on the $2$-BIS, as shown in Fig. \ref{fig:2D_BIS_a} (b-2).
We can see that the DSTs are anti-parallel on ${k_x} = 0$, while they are parallel on ${k_x} = \pi$. Accordingly, the former is nontrivial and the latter is trivial. On the whole, the $2$D system is topologically nontrivial with the topological invariant $\mathrm{T}_{2\text{-BIS}} = 1$.
In order to verify the correctness of the above result of the high-order BIS, we also investigate the DSTs on the $1$-BIS, as presented in Fig. \ref{fig:2D_BIS_a} (a-2). While the DST exhibits a non-zero $1$D winding along $k_{x} = 0$, there is no winding along $k_{x} = \pi$. The total non-zero winding number that characterizes the topologically nontrivial phase is consistent with that on the $0$D $2$-BIS.

\begin{figure}[tbp]
\centering
\setlength{\abovecaptionskip}{2pt}
\setlength{\belowcaptionskip}{4pt}
\includegraphics[angle=0, width=1 \columnwidth]{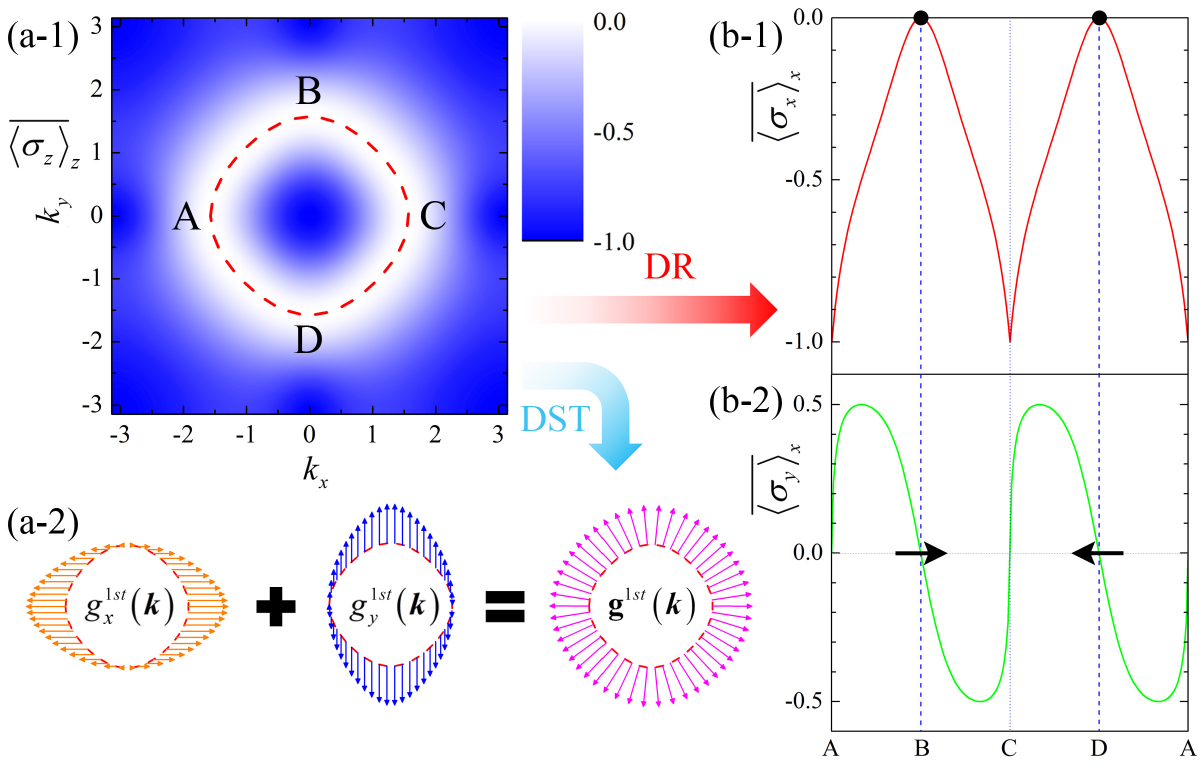}
\caption{(a-1) TASP ${\overline {\left\langle {{\sigma _z}} \right\rangle } _z}$ in the BZ after a quench from a large $m_x$ to $0$. The $1$-BIS is marked by a red dashed closed loop. (a-2) DSTs $g^{1st}_x$, $g^{1st}_y$, and $\mathbf{g}^{1st}$ on the $1$-BIS. (b-1) ${\overline {\left\langle {{\sigma _x}} \right\rangle } _x}$ and (b-2) ${\overline {\left\langle {{\sigma _y}} \right\rangle } _x}$ on the $1$-BIS after a quench from a large $m_x$ to $0$. Black points and arrows represent the $2$-BIS and DSTs $\mathbf{g}^{2nd}$, repectively.}
  \label{fig:2D_BIS_b}
\end{figure}

The validity of the present scheme can be confirmed with a different quench order. We first quench the $h_z$ axis and a new $1$D BIS, which is a closed loop, is obtained through measuring ${\overline {\left\langle {{\sigma _z}} \right\rangle } _z}$, as shown in Fig. \ref{fig:2D_BIS_b}.
Then on the BIS we quench the $h_x$ axis and measure ${\overline {\left\langle {{\sigma _x}} \right\rangle } _x}$. A $0$D BIS is formed by two $k$ points ($B$ and $D$) with ${\overline {\left\langle {{\sigma _x}} \right\rangle } _x} = 0$.
Through analyzing DSTs on the two BISs, we can conclude that the system is topologically nontrivial with $\mathrm{T}_{2\text{-BIS}} = 1$, in agreement with the former result.

\begin{figure*}[htbp]
\centering
\setlength{\abovecaptionskip}{2pt}
\setlength{\belowcaptionskip}{4pt}
\includegraphics[angle=0, width=0.8 \linewidth]{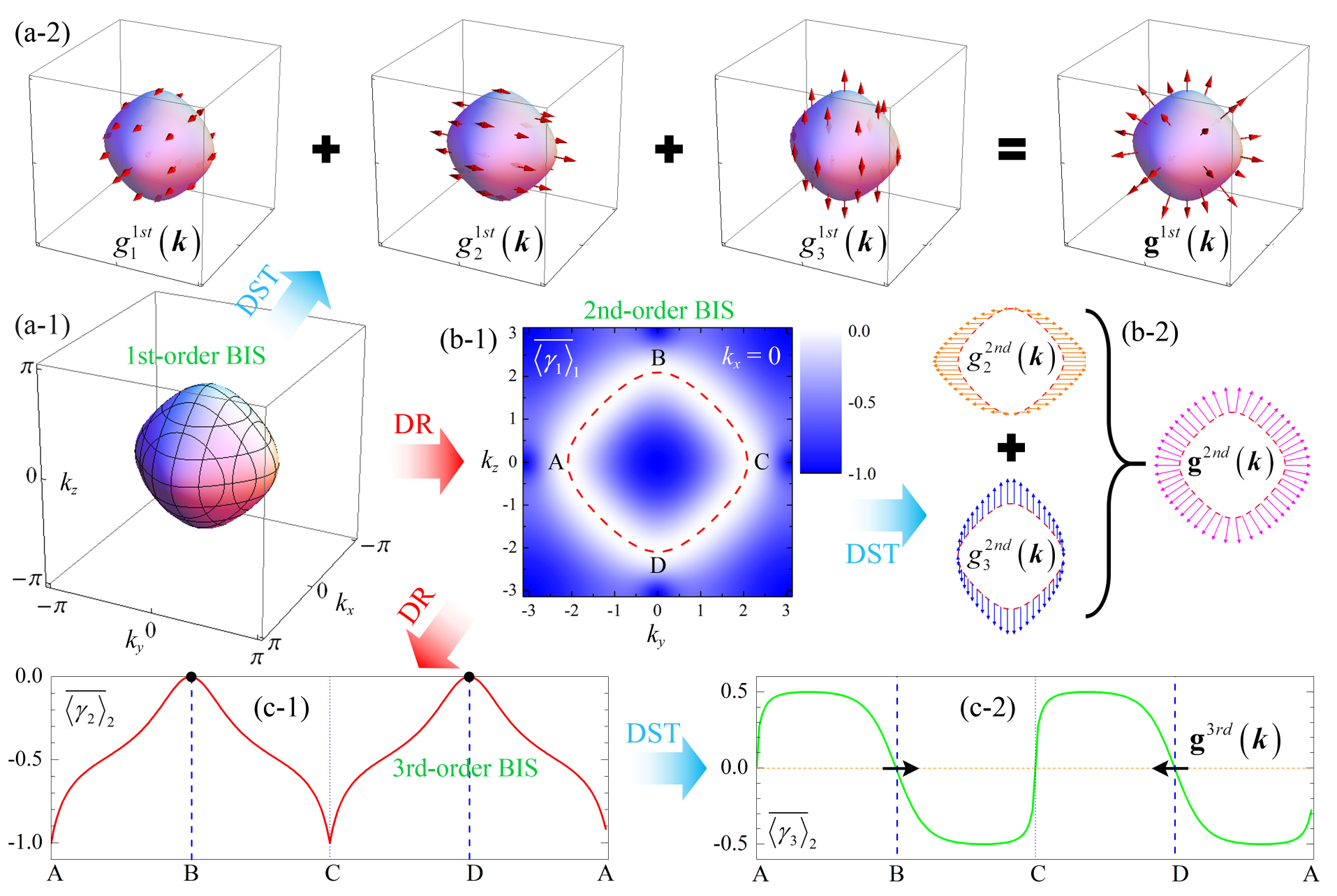}
\caption{
(a-1) The $1$-BIS with ${\overline {\left\langle {{\gamma _0}} \right\rangle } _0} = 0$ in the BZ.
(a-2) The DSTs $g^{1st}_1$, $g^{1st}_2$, $g^{1st}_3$, and $\mathbf{g}^{1st}$ on the $1$-BIS.
(b-1) The TASP ${\overline {\left\langle {{\gamma _1}} \right\rangle } _1}$ at the plane of $k_x = 0$. The $2$-BIS with ${\overline {\left\langle {{\gamma _1}} \right\rangle } _1} = 0$ is marked by a red dashed closed loop (A-B-C-D-A).
(b-2) The DSTs $g^{2nd}_2$, $g^{2nd}_3$, and $\mathbf{g}^{2nd}$ on the $2$-BIS.
(c-1) The TASP ${\overline {\left\langle {{\gamma _2}} \right\rangle } _2}$ on the $2$-BIS. The $3$-BIS is marked by two black points B and D.
(c-2) The TASP ${\overline {\left\langle {{\gamma _3}} \right\rangle } _2}$ on the $2$-BIS. Black arrows correspond to the DST $\mathbf{g}^{3rd}$ on the $3$-BIS. Red and blue arrows correspond to the dimension reduction (DR) of BISs and the calculation of DSTs, respectively.}
  \label{fig:3D_BIS}
\end{figure*}
We also simulate the dynamical scheme for $3$D system which we study experimentally later. The Hamiltonian
\begin{eqnarray}\label{eq:H3D}
{H_{3D}}{\rm{ = }}{h_0}{\gamma _0} + {h_1}{\gamma _1} + {h_2}{\gamma _2} + {h_3}{\gamma _3},
\end{eqnarray}
with ${h_0} = {m_0} - {t_0}\sum_i {\cos {k_{{i}}}} ,{h_i} = {m_i} + {t_{so}}\sin {k_{{i}}}$ for $i = 1, 2, 3$ (or $x, y, z$), ${t_{so}} = {t_0}$, $m_{0} = 1.5{t_0}$, $m_{1\sim 3}=0$, and $\gamma_{0 \sim 3}$ are Dirac matrices~\cite{note1}. We first quench the $h_0$ axis from a large $m_0$ to $m_0=1.5 t_0$, and measure TASP ${\overline {\left\langle {{\gamma _0}} \right\rangle } _0}$ in the whole BZ. This determines the $1$-BIS with $\overline {\left\langle {{\gamma _0}} \right\rangle } _0=0$, which is a closed surface, as plotted in Fig.~\ref{fig:3D_BIS} (a-1). We then quench the $h_1$ axis from a large $m_1$ to $m_1=0$, and measure ${\overline {\left\langle {{\gamma _1}} \right\rangle } _1}$. With this we obtain the $2$-BIS with $\overline {\left\langle {{\gamma _1}} \right\rangle } _1=0$ on the $1$-BIS, as shown in Fig.~\ref{fig:3D_BIS} (b-1). In the same way, we continue to quench the $h_2$ axis and measure ${\overline {\left\langle {{\gamma _2}} \right\rangle } _2}$ on the $2$-BIS.
The BIS is finally reduced to zero dimension, $3\text{-BIS}=\{\mathbf{k}\vert {\overline {\left\langle {{\gamma _0}} \right\rangle } _0}(\bold k)= {\overline {\left\langle {{\gamma _1}} \right\rangle } _1}(\bold k)={\overline {\left\langle {{\gamma _2}} \right\rangle } _2}(\bold k)=0\}$.
It consists of two momentum points (B and D) with $(k_x, k_y, k_z) = (0, 0, \pm \frac{2}{3} \pi)$, as marked in Fig.~\ref{fig:3D_BIS} (c-1). Around the two points we measure the TASP ${\overline {\left\langle {{\gamma _3}} \right\rangle } _2}$ with the same quench and calculate the corresponding DST $g^{3\rm rd}_3(\mathbf{k})$ on the $3$-BIS. In Fig.~\ref{fig:3D_BIS} (c-2) one can see that the DSTs are anti-parallel, giving topological invariant of post-quench $3$D system $\mathrm{T}_{3\text{-BIS}} = 1$. This result is clearly consistent with $\mathrm{T}_{1\text{-BIS}}$ ($\mathrm{T}_{2\text{-BIS}}$) obtained by the spin textures on the $2$D $1$-BIS ($1$D $2$-BIS), as presented in Figs.~\ref{fig:3D_BIS} (a-2) and (b-2).

The characterization with high-order dBSC necessitates far less measurements of spin evolutions than that with low-order dBSC. As pointed out, in determining the information of $n$-BIS, measurement is required only on the $(n-1)$-BIS, not in the whole BZ. This greatly simplifies the measurement strategy in real experiments.

\section{Dynamical characterization and simulation:~scheme II}
We proceed to study the second scheme to implement the dBSC and quantum simulation.
Unlike the above dynamical scheme, the second one necessitates to measure fewer or even only a single spin component to determine the complete topological information of the system, which further optimizes the experimental studies.

\begin{figure}[ht]
\centering
\setlength{\abovecaptionskip}{2pt}
\setlength{\belowcaptionskip}{4pt}
\includegraphics[angle=0, width=1 \linewidth]{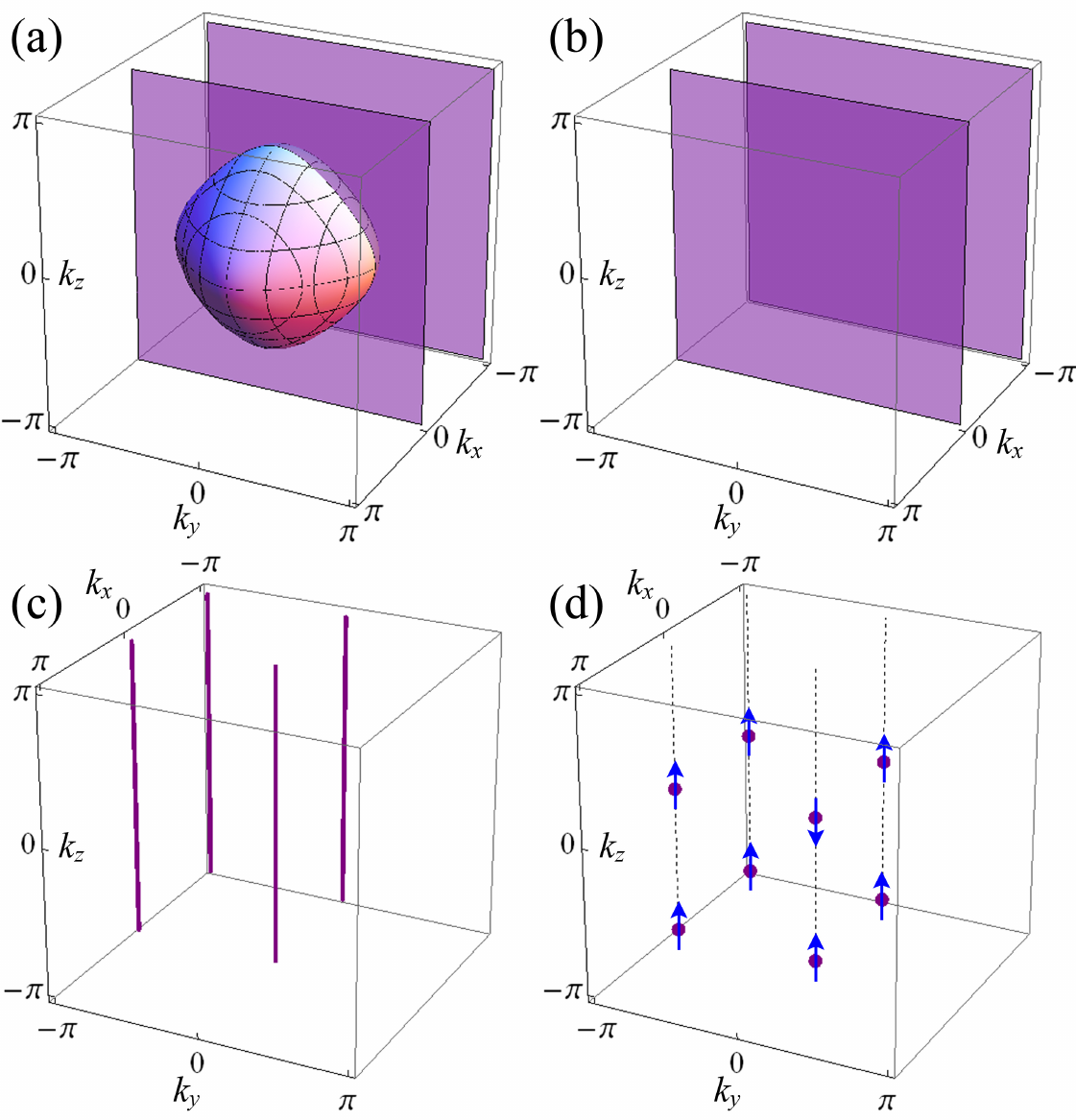}
\caption{(a) The BISs with ${\overline {\left\langle {{\gamma _0}} \right\rangle } _1} = 0$. (b) The $1$-BIS formed by two planes $k_x = 0$ and $- \pi$. (c) The $2$-BIS formed by four lines $(k_x, k_y)=\{(0,0),(0,-\pi),(-\pi,0),(-\pi,-\pi)\}$. (d) The $3$-BIS formed by eight points with $k_{x, y, z} = 0,-\pi$. Blue arrows represent the DST $g_0^{3rd}(\mathbf{k})$.}
  \label{fig:schemeII}
\end{figure}
We present the dynamical characterization with two steps. First, we determine the high-order BISs by measuring only the spin component $\gamma_0$. In particular, we perform the quench along the $\gamma_1$ axis and measure ${\overline {\left\langle {{\gamma _0}} \right\rangle } _1}$, which vanishes on two types of $1$-BIS. One is identical to that determined by ${\overline {\left\langle {{\gamma _0}} \right\rangle } _0}=0$, and another is identical to that by ${\overline {\left\langle {{\gamma _1}} \right\rangle } _1}=0$, as introduced in the characterization scheme I. We keep only the later one. Then we apply the quench along $\gamma_2$ axis and measure ${\overline {\left\langle {{\gamma _0}} \right\rangle } _2}$, which also leads to two types of $1$-BIS, respectively identical to ${\overline {\left\langle {{\gamma _0}} \right\rangle } _0}=0$ and ${\overline {\left\langle {{\gamma _2}} \right\rangle } _2}=0$. We again keep the later one, which intersects with the $1$-BIS identical to ${\overline {\left\langle {{\gamma _1}} \right\rangle } _1}$ and gives the $2$-BIS. With this dimension reduction process, we can obtain the general $n$th order BIS by $n\text{-BIS}= \{\mathbf{k}\vert {\overline {\left\langle {{\gamma _0}} \right\rangle } _i}=0, i = 1, 2, \dots, n\} - \{\mathbf{k}\vert {\overline {\left\langle {{\gamma _0}} \right\rangle } _0}=0\}$. Then, we measure the remaining spin components $\gamma_{l}$ to determine the DST as
\begin{eqnarray}\label{eq:g2}
g^{n\rm th}_{l}\left( \mathbf{k} \right) =  \frac{1}{{{\Upsilon _\mathbf{k}}}}\frac{{\partial {{\overline {\left\langle {{\gamma _l}} \right\rangle}_{n} }}}}{{\partial {k_ \bot }}}, \ l=0, n+1, n+2, \cdots, d.
\end{eqnarray}
Compared with Eq. (\ref{eq:g}), it shows qualitatively the same physics that is the variation of the remaining TASP components, including the $0$th one, across the $n$-BIS.
With the results of $\mathbf{g}^{n\rm th}(\mathbf{k})$ the topological invariant on the $n$-BIS is simply given by ${\mathrm T}_{n\text{-BIS}}[\mathbf{g}]$. In this characterization the $d-n+1$ components of spin-polarization need to be measured. Similarly, for the case $n=d$, we determine the topology by $g_{0}\left( \mathbf{k} \right)$ on the $d$-BIS,
${\mathrm T}_{d\text{-BIS}}[\mathbf{g}]
   = \frac{1}{2}\sum_{d\text{-BIS}_j} {\left[ {{\mathop{\rm sgn}} \left( {{g^{d\rm th}_{0,{L_j}}}} \right) - {\mathop{\rm sgn}} \left( {{g^{d\rm th}_{0,{R_j}}}} \right)} \right]}$,
corresponding to the coverage of the DST ${g}_0^{d\rm th}(\mathbf{k})$ over the 0D spherical surface $S^{0}$, i.e. the $0$D $d$-BIS.
Only the $\gamma_0$ component needs to be measured in determining the BISs and dynamical topological invariant in all $d$ times of quenches.

We present dynamical simulation with this approach for the $3$D system described in Eq.~\eqref{eq:H3D}, with only the $\gamma_0$-component being measured in each quench. Fig.~\ref{fig:schemeII} (a) shows the common $1$-BIS$_0$ with ${\overline {\left\langle {{\gamma _0}} \right\rangle } _0} = 0$, as well as the other one identical to ${\overline {\left\langle {{\gamma _1}} \right\rangle } _1} = 0$. They emerge in quenching the $h_1$ axis and measuring ${\overline {\left\langle {{\gamma _0}} \right\rangle } _1}$.
Keeping the latter yields the $1$-BIS, which are two planes with $k_x = 0$ and $\pi$ [Fig.~\ref{fig:schemeII} (b)].
Further, by respectively quenching $h_2$ and $h_3$ axes, and measuring ${\overline {\left\langle {{\gamma _0}} \right\rangle } _2}$ and ${\overline {\left\langle {{\gamma _0}} \right\rangle } _3}$ in the same way, we eventually obtain the $2$-BIS and $3$-BIS for the $3$D system. The result shows that the two BISs are respectively formed by four lines and eight points, as plotted in Figs.~\ref{fig:schemeII} (c) and (d). Finally, from the result of ${\overline {\left\langle {{\gamma_0}} \right\rangle } _3}$ around the $3$-BIS we obtain $g^{3\rm rd}_{0}\left( \mathbf{k} \right)$, which explicitly exhibit opposite values only in one pair of momenta with $k_x=k_y=0$ among the eight points, giving the topological invariant ${\mathrm T}_{3\text{-BIS}}[g^{3\rm rd}_{0}] = 1$. This is in agreement with the results shown in Fig.~\ref{fig:3D_BIS}, while the configurations of the BISs are very different due to the different quench and measurement sequences.

\textit{Advantages of characterization with high-order BISs}.--We make comments on the above dynamical schemes. For the characterization based on the $n$th-order BIS, the total number of measurements is $d+1$ to determine the BIS and corresponding topological invariant in the both dynamical schemes. Nevertheless, the measurement strategy is largely simplified (optimized) for characterization with higher-order (highest-order) BISs. We compare the two extreme cases with $n=1$ and $n=d$. In the former case one takes a single quench and measurement to determine $1$-BIS, but measures TASPs for the $d$ different spin components near the ($d-1$)D $1$-BIS to obtain the corresponding DST $\mathbf{g}^{1\rm st}(\mathbf{k})$. In the latter one takes $d$ times quenches to determine $d$-BIS, but measures TASPs along a single direction [within the $(d-1)$-BIS] near several discrete points of $d$-BIS, which is much simplified. Moreover essentially, during the dimension reduction for $n=d$ the number of momentum points necessitating measurement decreases very rapidly to extract the high-order BIS information, so that actual detection is remarkably simplified compared with that for $n=1$. For the second dynamical scheme, only the $\gamma_0$-component along a single direction is measured for $n=d$ to determine the dynamical topological invariant, which might be useful for real experiments. In general quantum simulation of the dynamical scheme with $d$-BISs is optimized.

\section{Experimental demonstration with quantum simulator}

\begin{figure*}[htbp]
\includegraphics[width=0.95 \linewidth]{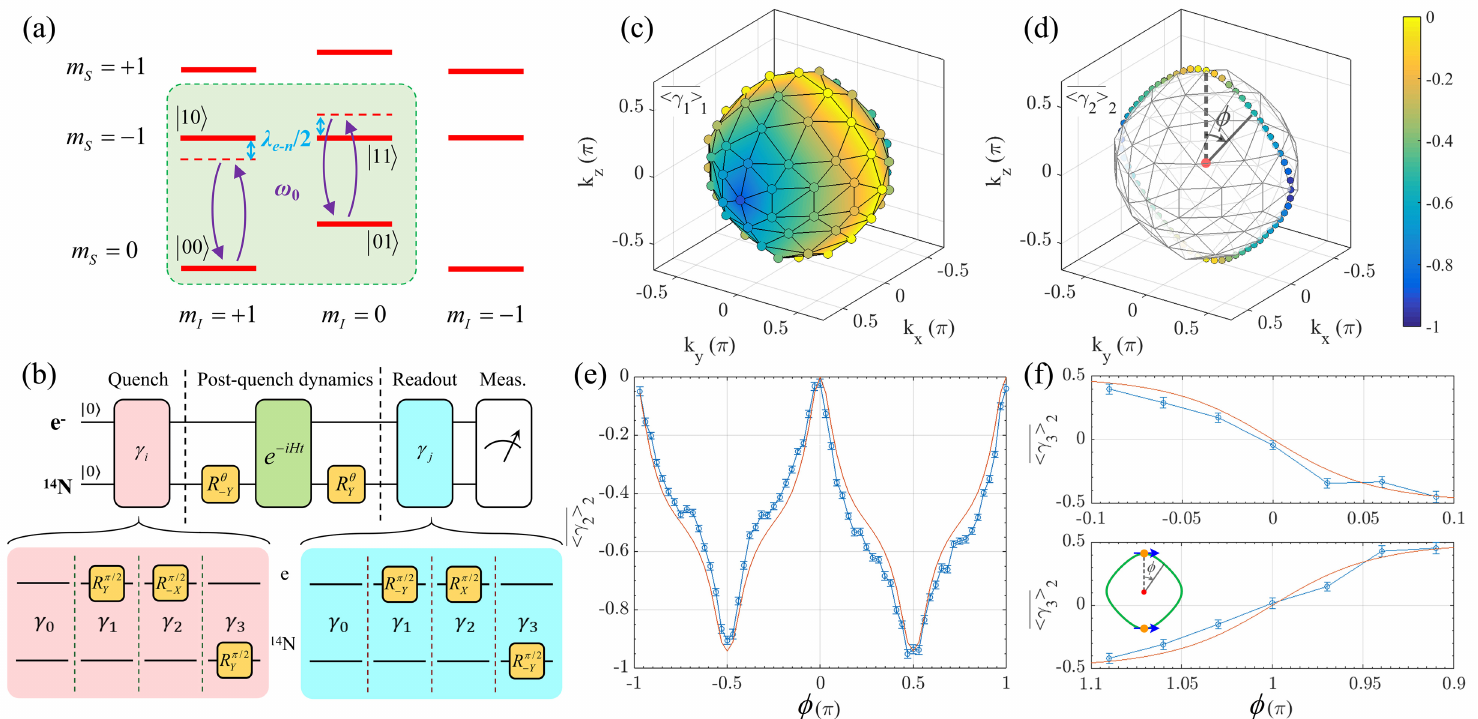}
\caption{(a) Energy levels of the NV center, the four in green frame are employed for the experiment and coupled by a microwave pulse $\omega_0$.
(b) Upper panel: quantum circuit for the dynamical simulation. The operator $e^{-iHt}$ denotes the evolution by $H_\text{RWA}$, while the net effect together with two nuclear spin operations $R^{\theta}_{\pm Y}$ renders an evolution under $H_\text{eff}$. Lower panel: operations of quench and read-out steps in different directions. (c-f) Experimental measurements of TASPs $\overline {\left\langle \gamma_1 \right\rangle }_1$ on the $1$-BIS, $\overline {\left\langle \gamma_2 \right\rangle }_2$ on the $2$-BIS, and $\overline {\left\langle \gamma_3 \right\rangle }_2$ near the $3$-BISs for $m_0 = 1.4t_0$ and $t_{so} = 0.2 t_{0}$. $\phi$ marked in (d) corresponds to the momentum on the $2$-BIS. Solid red lines in (e) and (f) are theoretical results for comparison. In the inset of (f), blue arrows represent the DST $g_3^{3rd}(\mathbf{k})$ on the $3$-BIS.}
\label{fig:experiment}
\end{figure*}

We present now the experimental study to verify the advantages of high-order BISs in topological characterization. As the study is carried out in momentum space, we can measure the high-order dynamical bulk-surface correspondence by emulating the BZ with one set of parameters for one momentum. We build the quantum simulator from solid-state spins of NV center in diamond~\cite{PR528.1,Ji2020} and simulate the Scheme I. A triplet ground state ($S =1$) of the electrons around the center and the intrinsic nitrogen-$14$ ($^{14}$N) nuclear spin ($I = 1$) form a coupled system [see Fig. \ref{fig:experiment} (a)]. The Hamiltonian reads
\begin{eqnarray}\label{exp1}
H_\text{NVC} = 2 \pi (\alpha_e \hat{s}_z + \beta_e \hat{s}_z^2 + \alpha_n \hat{i}_z +\beta_n \hat{i}_z^2 + \lambda_{e-n} \hat{s}_z \hat{i}_z),
\end{eqnarray}
where $\hat{s}_z$ and $\hat{i}_z$ respectively denote electron and nuclear spin operators.
A magnetic field of $446$ Gauss is applied along the NV's symmetry axis. It allows optical nuclear polarization~\cite{PRL102.057403} and leads to an electron Zeeman splitting $\alpha_e = 1250$ MHz and an nuclear one $\alpha_n = 137$ kHz. Besides, the other parameters $\beta_e = 2.87$ GHz, $\beta_n = -4.95$ MHz, and $\lambda_{e-n} = - 2.16$ MHz are the electronic zero-field splitting, the nuclear quadrupolar interaction ,and the hyperfine interaction, respectively.
The two-qubit subsystem is formed by the subspace of $\{m_S = 0, -1\} \otimes \{m_I = +1, 0\}$, relabeled as ${\left\{ {\left| 0 \right\rangle ,\left| 1 \right\rangle } \right\}_S} \otimes {\left\{ {\left| 0 \right\rangle ,\left| 1 \right\rangle } \right\}_I}$, on which the Pauli operators $\sigma$ and $\tau$ are defined.
We apply a microwave pulse to produce a driving field of strength $\omega_0$ and obtain an effective Hamiltonian $H_\text{RWA} = 2\pi (\frac{\lambda_{e-n}}{4}\sigma_{z}\otimes\tau_{z}+\omega_{0} \cos\varphi \sigma_x \otimes\mathds{1} - \omega_{0} \sin\varphi \sigma_y \otimes\mathds{1})$ under the rotating-wave approximation, where $\varphi$ is the phase of the microwave pulse.
A $\tau_y$-rotation of angle $\theta$ driven by a radio-frequency pulse transforms the Hamiltonian into
\begin{widetext}
\begin{eqnarray}\label{eq:Heff}
  H_\text{eff} = 2 \pi (\frac{\lambda_{e-n} \cos \theta}{4}\sigma_{z}\otimes\tau_{z}+\omega_{0} \cos\varphi \sigma_x \otimes\mathds{1} - \omega_{0} \sin\varphi \sigma_y \otimes\mathds{1} + \frac{\lambda_{e-n} \sin \theta}{4}\sigma_{z}\otimes\tau_{x}),
\end{eqnarray}
\end{widetext}
where the magnitudes of all the coefficients can be rescaled by an overall factor based on the evolution time (see details in Appendix B). From the above experimental operations, we can map the momentum-space parameters $\mathbf{k}$ to the experimental parameters $(\omega_{0}, \varphi, \theta)$ and successfully emulate with $H_\text{eff}$ the $3$D chiral topological insulator given by the Hamiltonian $H_{3D}$ [Eq. (\ref{eq:H3D})].

We measure the quench dynamics to determine the bulk topology. The quantum circuit for the simulation experiment is plotted in Fig. \ref{fig:experiment} (b) and simulated parameters are set as $m_0 = 1.4t_0$ and $t_{so} = 0.2 t_0$ in Eq. (\ref{eq:H3D}). There are three main steps in the whole experimental procedure.
(i) We prepare an initial state $\left|00\right\rangle$ and rotate the polarization direction to ``$0$" by a unitary control. Then, quench $m_0$ and the state evolves under the Hamiltonian $H_\text{eff}$, emulating $H_{3D}$. Meanwhile, measure the TASP ${\overline {\left\langle {{\gamma _0}} \right\rangle }_0}$ in the BZ to obtain the $1$-BIS, which is a closed surface~\cite{Ji2020}.
(ii) Repeat the first step to obtain the $2$-BIS and the $3$-BIS through quenching and measuring along ``$1$" and ``$2$" directions, but the measurements are only taken on the $1$-BIS and the $2$-BIS, respectively, much less than whole BZ. In Fig.~\ref{fig:experiment} (c-e) we can see that the former with ${\overline {\left\langle {{\gamma _1}} \right\rangle }_1} = 0$ is a closed loop in the $k_x = 0$ plane and the latter with ${\overline {\left\langle {{\gamma _2}} \right\rangle }_2} = 0$ includes only two $\mathbf{k}$ points.
(iii) The topological invariant is finally determined by the DST $g_3^{3rd}(\mathbf{k})$ given from the variation of ${\overline {\left\langle {{\gamma _3}} \right\rangle }_2}$ across the two points of the $3$-BIS [Eq.~(\ref{eq:g})]. The experimental results are in good agreement with theoretical calculations in Fig. \ref{fig:experiment} (f) and the simulated system is topologically nontrivial due to the anti-parallel DSTs on the $1$D 2-BIS. Compared with previous studies where one needs to measure TASPs for three different spin components near the $2$D $1$-BIS to obtain the corresponding DST~\cite{Xin2020,Ji2020}, which contains a large number of momentum points, here one only needs to measure TASPs along a single direction near two single points of $3$-BIS.
Even though we also need another step to reduce the dimension of BISs, the total number of actual measurements is much less and thus the strategy is greatly simplified.

The above experimental study verifies that the characterization and simulation of topological phases using high-order BISs are of clear advantages and high efficiency. While this verification does not reply on the construction of a real topological band for the single-particle regime, the spin-qubit simulator can be further applied to investigate the interacting effects if taking into account more qubits. The interacting effects on the quench dynamics can be simulated by simultaneous control of the different spin qubits during the evolution.
The topological characterization based on high-order BISs could provide optimal schemes to explore the correlated topological quench dynamics in experiment, which shall be considered in our next work.

\section{Conclusion and outlook}

We have proposed the concept of high-order BISs, with which we developed a new dynamical theory to characterize and simulate topological quantum phases, and experimentally built up a quantum simulator using spin qubits to verify the feasibility of the new theory in the dynamical characterization through emulating different momenta one by one.
Especially, for the highest-order BIS with zero dimension, the topological invariants can be determined with the minimal measurement strategy, showing the optimal scheme with fundamental advantages in characterizing and simulating topological phases, as verified in our simulation experiment.

Theoretically, the present dimension reduction through high-order BISs is performed by reducing the degree of freedom corresponding to the Clifford algebra space, which is similar to, but different from that in typical classification theories like $K$-theory~\cite{Ktheory} for the topological phases. The dimension reduction in the $K$-theory is performed directly in the real or momentum space. While so far the dynamical theory is developed for topological phases with integer invariants, the dimension reduction approach proposed here shows important insight into the feasibility of establishing the dynamical classification of the complete set of topological phases~\cite{PRB78.195125,Chiu2016,Kitaev2009} in the AZ ten-fold symmetry classes~\cite{PRB55.1142} and further crystalline topological phases~\cite{Chiu2016}, which is stimulating and shall be presented in our next works.

\begin{acknowledgments}
We thank Wei Sun for helpful discussion. This work was supported by National Natural Science Foundation of China (Grant Nos. 11674152, 11825401, 11761161003, U1801661, and No. 11775209), Guangdong Innovative and Entrepreneurial Research Team Program (Grant No. 2016ZT06D348), Guangdong Provincial Key Laboratory (Grant No. 2019B121203002), Science, Technology, and Innovation Commission of Shenzhen Municipality (Grant No. JCYJ20190809120203655), the Open Project of Shenzhen Institute of Quantum Science and Engineering (Grant No.SIQSE202003), and the Fundamental Research Funds for the Central Universities.
\end{acknowledgments}

\appendix
\begin{figure*}[tbp]
\centering
\setlength{\abovecaptionskip}{2pt}
\setlength{\belowcaptionskip}{4pt}
\includegraphics[angle=0, width=0.95 \linewidth]{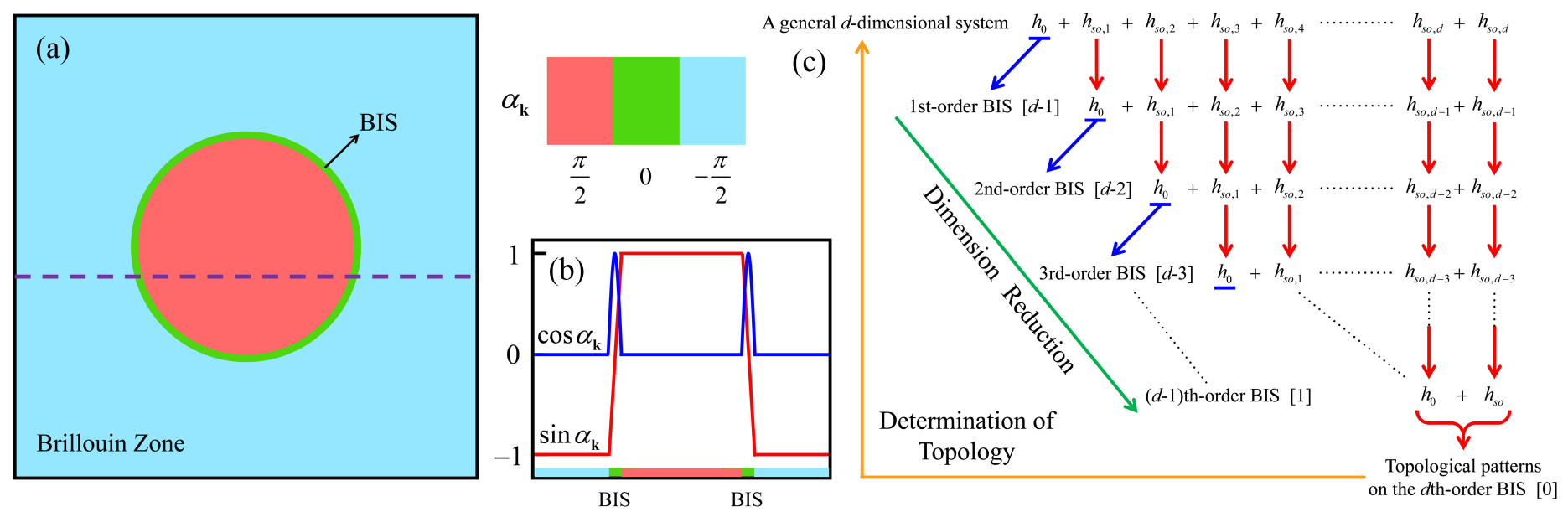}
\caption{(a) A sketch of the step function $\alpha_{\mathbf{k}}$ in the BZ. (b) The functions $\sin {\alpha_{\mathbf{k}}}$ (red) and $\cos {\alpha_{\mathbf{k}}}$ (blue) along the dashed line in (a). (c) A schematic diagram of the dimension reduction. The number in the square bracket is the dimensionality of the corresponding BIS.}
  \label{fig:alphak_BIS}
\end{figure*}

\begin{figure*}[htbp]
\centering
\setlength{\abovecaptionskip}{2pt}
\setlength{\belowcaptionskip}{4pt}
\includegraphics[width=0.8 \linewidth]{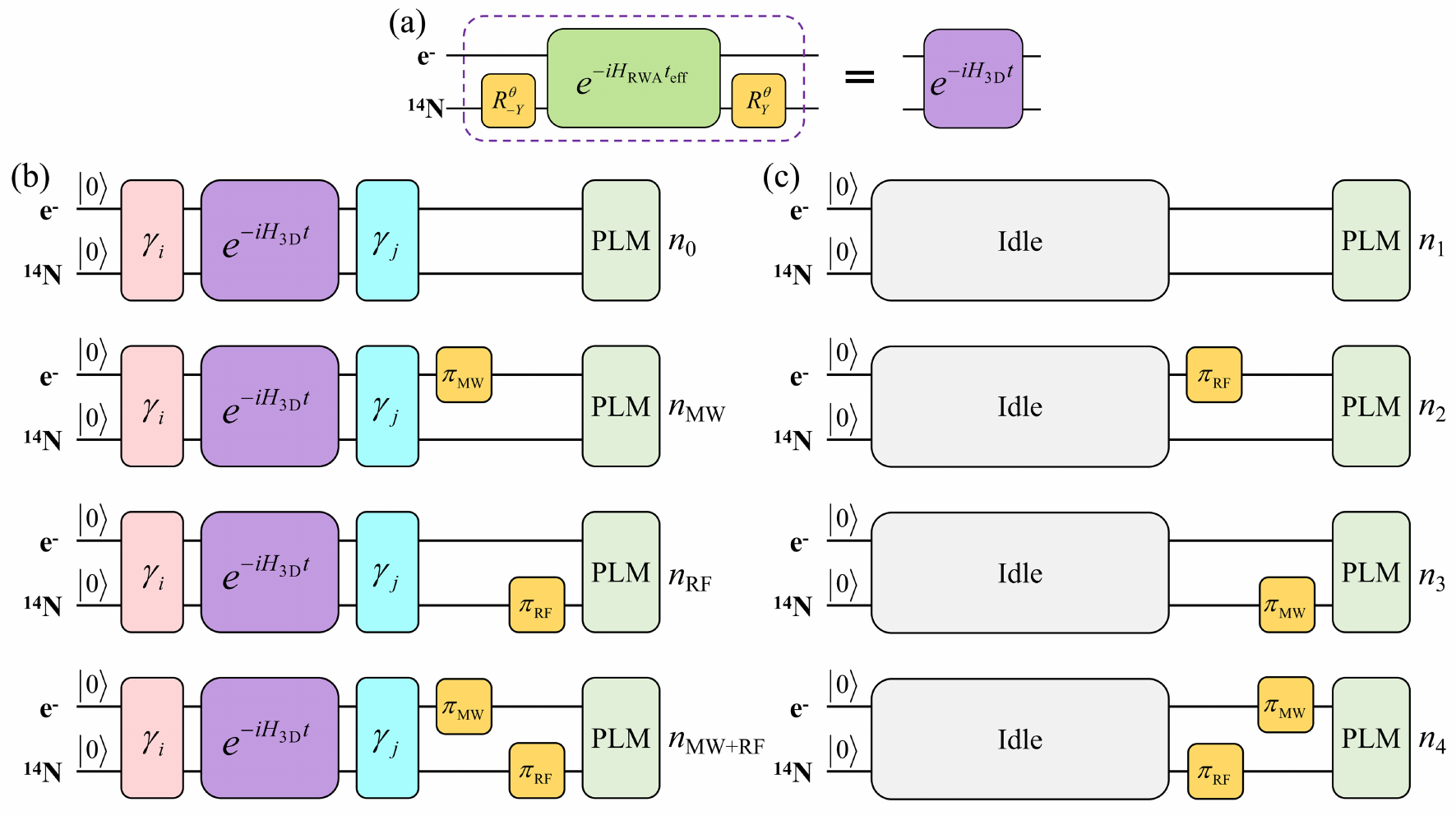}
\caption{
(a) Quantum circuit for post-quench dynamics on the left-hand side. $R^{\theta}_{\pm Y}$ represents rotation of nuclear spin about $\pm y$ axis by an angle $\theta$. $e^{-iHt}$ corresponds to evolution of the system under the rotating-wave approximation for a time duration of $t_\text{eff}$. The net effect is equivalent to the evolution under $H_{3D}$ for a duration of $t$ on the right-hand side.
(b-c) Sequences to measure populations. The operation labeled $\gamma_i$ (pink) is a deep quench along $i$ direction, which requires unitary operations to transform the state $|00>$ to an eigenstate of the pre-quench Hamiltonian. $e^{-iH_{3D}t}$ (purple) corresponds to the post-quench dynamical evolution, whoes detailed quantum cirtuit is plotted in (a). The operation $\gamma_j$ (azure) corresponds to measurement of the $\gamma_j$ component, which requires an operation to transform $\gamma_j$ to the $z$ basis.
$\pi_\text{MW}$ and $\pi_\text{RF}$ respectively  represent the microwave and ratio-frequency pulses, which are applied to rotate spins.
PLM (green) corresponds to a photoluminescence measurement. Idle corresponds to a waiting time equal to the total time of the $\gamma_i$, $H_{3D}$ and $\gamma_j$ steps.
}
\label{fig:quantum circuit}
\end{figure*}

\begin{figure}[tbp]
\centering
\setlength{\abovecaptionskip}{2pt}
\setlength{\belowcaptionskip}{4pt}
\includegraphics[angle=0, width=1 \columnwidth]{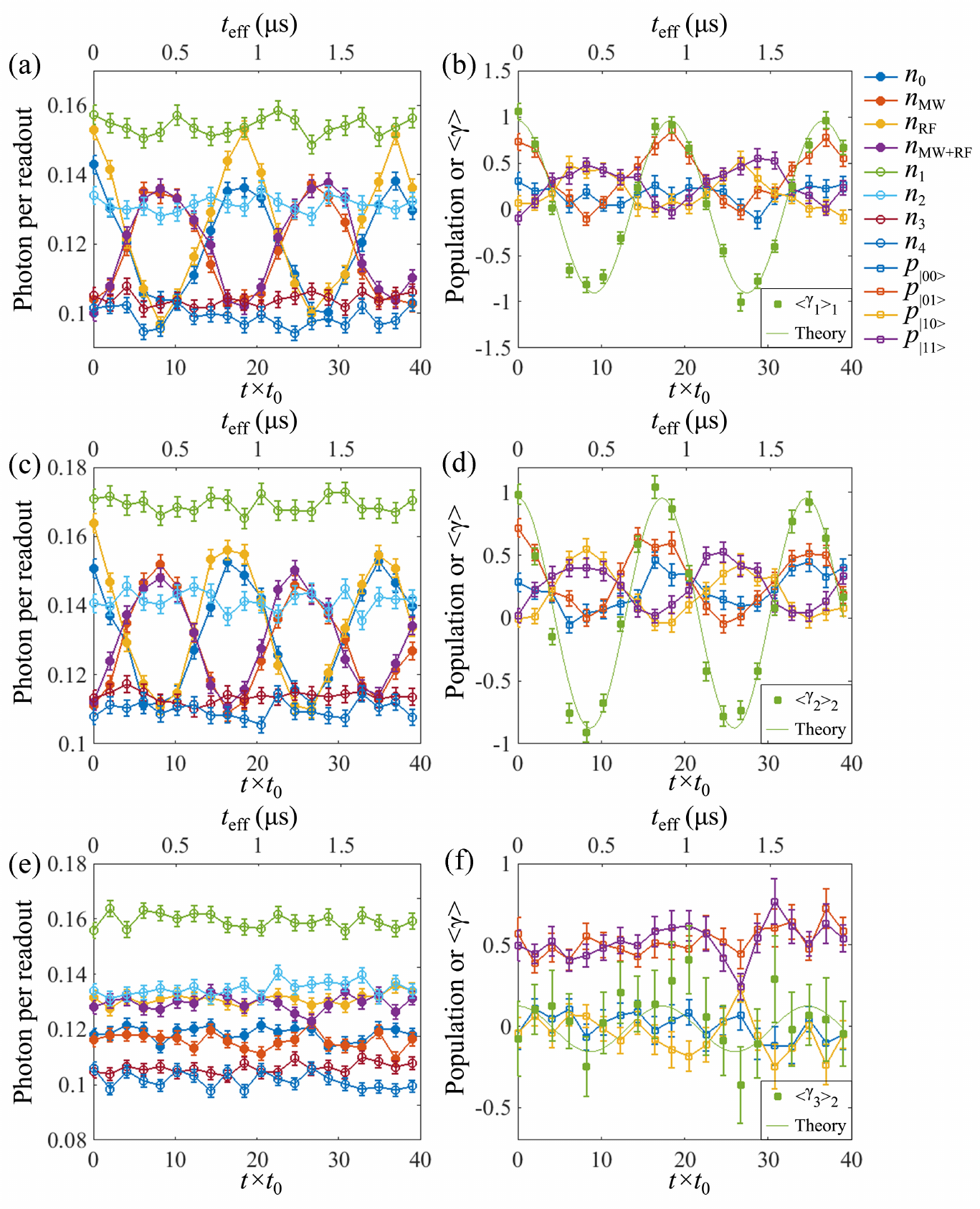}
\caption{
(a,b), (c,d), and (e,f) respectively show measurement data of ${\langle \gamma_1 \rangle}_1$, ${\langle \gamma_2 \rangle}_2$, and ${\langle \gamma_3 \rangle}_2$ on the $3$-BIS. (a), (c), and (e) plot average photon number for different sequences as shown in Fig. \ref{fig:quantum circuit} at different time. Error bars are estimated considering photon shot noise. In (b), (d), and (f) populations and spin polarizations are obtained from results in (a), (c), and (e), respectively. Solid green lines in (b), (d), and (f) are calculated from the theoretical model. We set $m_0 = 1.4t_0$, $t_{so} = 0.2t_0$, and the momentum point on the $3$-BIS corresponds to $\phi = \pi$ in Fig. \ref{fig:experiment}.
}
\label{fig:experimental data}
\end{figure}

\section{High-order bulk-surface duality}

Topological invariants of band insulators are usually defined in the whole or half BZ~\cite{RMP82.3045,RMP83.1057,RMP88.035005}. The bulk-surface duality~\cite{SB63.1385} shows that the integer topological invariant of a $d$-dimensional ($d$D) topological phase can be generically mapped to the topological number defined on the ($d-1$)D BIS~\cite{SB63.1385}.
Here we propose the concept of high-order BISs by introducing a dimension reduction approach, and show that the topological invariant of a $d$D topological phase can be reduced to the lower-dimensional topological numbers on arbitrary high-order BISs, rendering the high-order bulk-surface duality.

We start from a general $d$D gapped phase within the $Z$ classification. Its corresponding Hamiltonian is given in Eq. (\ref{eq:general H}).
Here, for the convenience of following description, we choose any component of the Hamiltonian as ${h_0}\left( \mathbf{k} \right)$ and the other components can be regarded as a spin-orbit field ${h_{so,i}}\left( \mathbf{k} \right)$ with $i = 1 \sim d$.
As well known, the topology of a system remains unchanged under any continuous deformation that does not close the energy gap. Thus, we can perform a general analysis and only consider the spin-orbit field near the BIS with zero field in the other region. For this purpose, we transform the Hamiltonian into such a form,
\begin{eqnarray}\label{eq:deformed_H}
  H\left( \mathbf{k} \right) & \Rightarrow & \mathbf{h} '\left( \mathbf{k} \right) \cdot \mathbf{\gamma}   = \sum\limits_{i = 0}^d {h{'_i}\left( \mathbf{k} \right){\gamma _i}} \nonumber \\
  &=& \sin {\alpha_{\mathbf{k}}}{\gamma _0} + \cos {\alpha_{\mathbf{k}}}\sum\limits_{i = 1}^d {{{\hat h}_{so,i}}\left( \mathbf{k} \right){\gamma _i}},
\end{eqnarray}
where ${\hat h_{so,i}}\left( \mathbf{k} \right) = {h_{so,i}}\left( \mathbf{k} \right)/\sqrt{\sum\limits_{j = 1}^d {h_{so,j}^2\left( \mathbf{k} \right)}}$ makes the Hamiltonian be normalized and ${\alpha_{\mathbf{k}}}$ is a step function, as illustrated in Fig. \ref{fig:alphak_BIS} (a). ${\alpha_{\mathbf{k}}} = \frac{\pi }{2}$, $-\frac{\pi }{2}$, and $0$, when $k$ points are inside, outside, and on the BIS, respectively.
For the $Z$ classification, when the system is odd (even)-dimensional, the corresponding topological invariant, uniformly defined as ${\mathrm T}_d$, is winding number (Chern number)~\cite{PRB88.075142,PRB90.165114,RMP88.035005}. Before performing a continuous dimension reduction of the BIS, we first investigate and calculate the topological invariants in general odd and even systems, respectively.

For a ($2n+1$)D system, its topological invariant is the winding number, defined as
\begin{equation}\label{}
  {{\mathrm T}_{2n + 1}} = \frac{{{{\left( { - 1} \right)}^n}n!}}{{2{{\left( {2\pi \mathbf{i}} \right)}^{n + 1}}\left( {2n + 1} \right)!}}\int_\text{BZ} {\mathrm{Tr}\left[ {\gamma H{{\left( {dH} \right)}^{2n + 1}}} \right]},
\end{equation}
where the Hamiltonian has a chiral symmetry $\gamma = \mathbf{i}^{n+1}\prod\limits_{i = 0}^{2n+1} \gamma_i$ and
\begin{eqnarray}
&&\textrm{Tr}\left[ {\gamma H{{\left( {dH} \right)}^{2n + 1}}} \right] \\
&=& \left( {2n + 1} \right)!{\left( { - 2 \mathbf{i}} \right)^{n + 1}}\sum\limits_{i = 0}^{2n + 1} {{\varepsilon ^{i01 \cdots \left( {2n+1} \right)}}h_i'\overline\bigwedge_{l = 0}^{2n+1}dh_l'}. \nonumber
\end{eqnarray}
Here $\overline\bigwedge_{l = 0}^{2n+1}dh_l' = dh_0' \wedge  \cdots  \wedge dh_{2n + 1}'$ with $l \ne i$.
The winding number can be simplified as
\begin{eqnarray}\label{eq:T_2n+1_BZ}
{{\rm{T}}_{2n+1}} = - \frac{{n!}}{{2{\pi ^{n+1}}}}{\int_{\text{BZ}} {\sum\limits_{i = 0}^{2n + 1} {
{\varepsilon ^{i01 \cdots \left( {2n + 1} \right)}}{{h}_{i}'}\overline\bigwedge_{l = 0}^{2n+1}d{{\hat h}_{l}'}}}}.
\end{eqnarray}
Substitute the above deformed Hamiltonian Eq. (\ref{eq:deformed_H}) into the wedge product term,
\begin{eqnarray}\label{eq:combination}
&&\int_{\text{BZ}} \sum\limits_{i = 0}^{2n+1} {{\varepsilon ^{i01 \cdots \left( {2n+1} \right)}}h_i'\overline  \bigwedge  _{l = 0}^{2n+1}dh_l'} \\
&=& \int_{\frac{\pi }{2}}^{ - \frac{\pi }{2}} {{{\cos }^{2n}}{\Theta _{k_{\bot} }}d{\Theta _{k_{\bot} }}} \sum\limits_j {\int_{\text{BIS}_j} {\sum\limits_{i = 1}^{2n+1} {{{\left( { - 1} \right)}^i}{{\hat h}_{so,i}}\overline \bigwedge_{l = 1}^{2n+1}d{{\hat h}_{so,l}}} } }, \nonumber
\end{eqnarray}
where $k_{\bot}$ denotes the momentum perpendicular to the BIS, $j$ corresponds to different sectors of the BIS, and
\begin{eqnarray}\label{}
&& \int_{\frac{\pi }{2}}^{ - \frac{\pi }{2}} {{{\cos }^{2n}}{\Theta _{k_{\bot} }}d{\Theta _{k_{\bot} }}}
 =  - \frac{{\left( {2n - 1} \right)!!}}{{{2^{n}}n!}}\pi. \nonumber
\end{eqnarray}
Finally, the winding number of the ($2n+1$)D system, defined on the $2n$D BIS, reads
\begin{eqnarray}\label{eq:T_2n+1_BIS}
{\mathrm T_{2n + 1}} = && - \frac{{\left( {2n - 1} \right)!!}}{{{2^{n + 1}}{\pi ^n}}}\times\\
&&\sum\limits_j {\int_{\text{BIS}_j} {\sum\limits_{i = 0}^{2n} {{\varepsilon ^{i01 \cdots \left( {2n} \right)}}{{\hat h}_{so,i}}\overline\bigwedge_{l = 0}^{2n}d{{\hat h}_{so,l}}}}}.\nonumber
\end{eqnarray}
We have changed the range of the index $i$ from $1 \sim 2n+1$ to $0 \sim 2n$.
One can see that the integrated area of the winding number has been changed from the ($2n+1$)D BZ to the $2n$D BIS. It manifests a bulk-surface correspondence.

For a $2n$D system, its topological invariant is the $n$th-order Chern number, defined as
\begin{equation}\label{eq:Chern number}
  {{\mathrm T}_{2n}} =  - \frac{1}{{{2^{2n + 1}}}}\frac{1}{{n!}}{\left( {\frac{\mathbf{i}}{{2\pi }}} \right)^n}\int_\text{BZ} {\mathrm{Tr}\left[ {H{{\left( {dH} \right)}^{2n}}} \right]},
\end{equation}
where
\begin{eqnarray}\label{eq:C_Tr}
\mathrm{Tr}\left[ {H{{\left( {dH} \right)}^{2n}}} \right] = \left( {2n} \right)!{\left( { - 2 \mathbf{i}} \right)^n}
\sum\limits_{i = 0}^{2n} {{\varepsilon ^{i01 \cdots \left( {2n} \right)}}h_i'\overline\bigwedge_{l = 0}^{2n}d{{h}_{l}'}}.
\end{eqnarray}
Substitute Eq. (\ref{eq:C_Tr}) into Eq. (\ref{eq:Chern number}),
\begin{eqnarray}\label{eq:T_2n_BZ}
{{\mathrm T}_{2n}}=- \frac{{\left( {2n - 1} \right)!!}}{{{2^{n + 1}}{\pi ^n}}}
\int_\text{BZ}{\sum\limits_{i = 0}^{2n} {{\varepsilon ^{i01 \cdots \left( {2n} \right)}}h_i'\overline\bigwedge_{l = 0}^{2n}dh_{l}'}}.
\end{eqnarray}
The Chern number can be calculated in the same way as the winding number. We can obtain
\begin{eqnarray}\label{eq:T_2n-1_BIS}
{{\rm{T}}_{2n}} = &&  - \frac{{\left( {n - 1} \right)!}}{{2{\pi ^n}}}\times \\
&&\sum\limits_j {\int_{\text{BIS}_j} {\sum\limits_{i = 0}^{2n - 1} {
{\varepsilon ^{i01 \cdots \left( {2n - 1} \right)}}{{\hat h}_{so,i}}\overline\bigwedge_{l = 0}^{2n-1}d{{\hat h}_{so,l}}}}}.\nonumber
\end{eqnarray}
Similarly, its integrated area is also changed from the $2n$D BZ to the ($2n-1$)D BIS with reducing one dimension.

Interestingly, through comparing the right-hand sides of Eqs. (\ref{eq:T_2n+1_BIS}) and (\ref{eq:T_2n_BZ}), which are respectively the winding number defined on the BIS of the ($2n+1$)D system and the Chern number defined in the BZ of the $2n$D system, we find that they have a similar form.
(i) Both of them have the same constant factor. (ii) In the former the summation over $i$ includes $2n+1$ components of the ($2n+1$)D system, while in the latter it includes all the components of the $2n$D system. It is important that the number of the components is equal to each other. (iii) In the former the integrated area is the BIS which are $2n$D, while in the latter it is the BZ which is also $2n$D.
Therefore, when the BIS of the ($2n+1$)D system is regarded as the BZ of a new $2n$D system, the two equations are completely equivalent.

Similarly, we compare Eq. (\ref{eq:T_2n-1_BIS}) and the winding number of a generic ($2n-1$)D system, which can be obtained from Eq. (\ref{eq:T_2n+1_BZ}) and reads
\begin{eqnarray}\label{}
{{\mathrm T}_{2n - 1}} = - \frac{{\left( {n - 1} \right)!}}{{2{\pi ^n}}}\times {\int_{\text{BZ}} {\sum\limits_{i = 0}^{2n - 1} {
{\varepsilon ^{i01 \cdots \left( {2n - 1} \right)}}{{h}_{i}'}\overline\bigwedge_{l = 0}^{2n-1}d{{\hat h}_{l}'}}}}.
\end{eqnarray}
We can see that they are also equivalent.

Based on the above analysis and calculations of general systems, a continuous dimension reduction of the BIS can be performed, as illustrated in Fig. \ref{fig:alphak_BIS} (c). In each process of the dimension reduction, we regard the obtained BIS as a new BZ for next dimension reduction. Repeating this process, the BIS can be reduced to ($d-n$) dimensions for a $d$D system. It is called the $n$th-order BIS, defined as
\begin{equation}
n\text{-BIS} = \{\mathbf{k}\in\mathrm{BZ}\vert h_{i_\alpha}(\mathbf{k})=0, \alpha=0,1,\dots,n-1\},
\end{equation}
where $h_{i_\alpha}$'s are $n$ arbitrary components chosen as $h_0$ during the dimension reduction.
The corresponding topological invariant, which is equal to the original invariant ${\mathrm T}_{d}$, is given as
\begin{equation}
{\mathrm T}_{n\text{-BIS}}=\frac{\Gamma[(d-n+1)/2]}{2\pi^{(d-n+1)/2}}\frac{1}{(d-n)!}\int_{n\text{-BIS}}
\mathbf{\hat{h}}_{so}(\mathrm{d}\hat{\mathbf{h}}_{so})^{d-n},
\end{equation}
where $\Gamma(a)$ is the Gamma function with $\Gamma(a+1) = a \Gamma(a)$, $\Gamma(\frac{1}{2}) = \sqrt{\pi}$, and $\Gamma(1) = 1$.
Finally, the BIS can be reduced to zero dimension and only consists of several $k$ points. The corresponding topological invariant is also greatly simplified [Eq. (\ref{eq:T_0_BIS})].
Obviously, the $d$-BIS is formed by the cross points of $d$ components of $\mathbf{h}\left( \mathbf{k} \right)$. They satisfy $\sum\limits_{m \ne i} {\left| {{h_m}\left( \mathbf{k} \right)} \right|}  = 0$, where ${h_i}$ is the only component that is not chosen as ${h_0}$ in Fig. \ref{fig:alphak_BIS} (c).

\section{Experimental simulation}
\subsection{Quench process $\And$ post-quench dynamics}
In our quantum simulator built up from the diamond NV center, we choose the subspace of $\left\{ {{m_S} = 0,\; - 1} \right\} \otimes \left\{ {{m_I} =  + 1,\;0} \right\}$  to perform the quantum simulation. The subspace Hamiltonian is in a diagonal form,
\begin{equation}\label{}
  {H_0} = 2\pi \left( {\begin{array}{*{20}{c}}
{\Omega_1}&0&0&0\\
0&{\Omega_2}&0&0\\
0&0&{\Omega_3}&0\\
0&0&0&{\Omega_4}
\end{array}} \right),
\end{equation}
where $\Omega_1 = {{\alpha _n} + {\beta _n}}$, $\Omega_2 = 0$, $\Omega_3 = { - {\alpha _e} + {\beta _e} + {\alpha _n} + {\beta _n} - {\lambda _{e - n}}}$, and $\Omega_4 = { - {\alpha _e} + {\beta _e}}$.

To simulate $\gamma_1$ and $\gamma_2$ terms in $H_{3D}$ [Eq. (\ref{eq:H3D})], a microwave pulse of frequency ${\omega _0} =  - {\alpha _e} + {\beta _e} - {\lambda _{e - n}}/2$ is applied to couple basis states, including $\left| {00} \right\rangle  \leftrightarrow \left| {10} \right\rangle$ and $\left| {01} \right\rangle  \leftrightarrow \left| {11} \right\rangle$. Its corresponding interacting Hamiltonian reads
\begin{equation}\label{}
  {H_\text{int}} = 2\pi {\omega _0}\cos \left( {{\omega _0}t + \varphi } \right)\left( {\begin{array}{*{20}{c}}
0&0&1&0\\
0&0&0&1\\
1&0&0&0\\
0&1&0&0
\end{array}} \right).
\end{equation}
Then we transform the total Hamiltonian ${H_\text{tot}} = {H_0} + {H_\text{int}}$ to the rotating frame defined by the microwave field. Under proper rotating-wave approximation the system Hamiltonian can be written as
\begin{equation}\label{}
{H_\text{RWA}} = 2\pi \left( {\frac{{{\lambda _{e - n}}}}{4}{\sigma _z}{\tau _z} + {\omega _0}\cos \varphi {\sigma _x} \otimes 1 - {\omega _0}\sin \varphi {\sigma _y} \otimes 1} \right).
\end{equation}

We further apply a rotation ${U_{rot}} = {e^{ - i\theta {\tau _y}/2}}$ to the system Hamiltonian, and then obtain the effective Hamiltonian Eq. (\ref{eq:Heff}).
At the same time, we imply $\theta = \arctan(h_3/h_1)$, and define the effective time as a rescale of the simulation time $t$, i.e., ${t_\text{eff}} = \kappa t$.
Our purpose is to reproduce the same effect as $U_{3D} = e^{-iH_{3D}t}$ with the simulated evolution $U_\text{eff} = e^{-iH_\text{eff} t_\text{eff}}$. Thus, $H_{3D} = \kappa H_\text{eff}$. The parameters satisfy
\begin{eqnarray}\label{}
  \kappa  &=& \frac{2}{{\pi \left| {{\lambda _{e - n}}} \right|}}\sqrt {h_0^2 + h_3^2} ,\nonumber \\
  {\omega _0} &=& \frac{1}{{2\pi \kappa }}\sqrt {h_1^2 + h_2^2} , \\
  \varphi  &=&  - \arctan \left( {{h_2}/{h_1}} \right). \nonumber
\end{eqnarray}

To show the mapping relation between the momentum space parameters and experimental ones, we substitute $h_{0,1,2,3}$ in the above equations with momentum space parameters, giving
\begin{equation}\label{}
\begin{array}{l}
\theta  = \arctan \frac{{\left| {{t_{so}}\sin {k_z}} \right|}}{{{m_0} - {t_0}\left( {\cos {k_x} + \cos {k_y} + \cos {k_z}} \right)}},\\
{\omega _0} = \frac{{\left| {{\lambda _{e - n}}} \right|}}{4}\sqrt {\frac{{t_{so}^2\left( {{{\sin }^2}{k_x} + {{\sin }^2}{k_y}} \right)}}{{{{\left[ {{m_0} - {t_0}\left( {\cos {k_x} + \cos {k_y} + \cos {k_z}} \right)} \right]}^2} + t_{so}^2{{\sin }^2}{k_z}}}} ,\\
\varphi  = \arctan \frac{{\sin {k_y}}}{{\sin {k_x}}},\\
{t_{{\text{eff}}}} = \frac{{2t}}{{\pi \left| {{\lambda _{e - n}}} \right|}}\sqrt {{{\left[ \begin{array}{l}
{m_0} - {t_0}\cos {k_x}\\
 - {t_0}\cos {k_y} - {t_0}\cos {k_z}
\end{array} \right]}^2} + t_{so}^2{{\sin }^2}{k_z}}.
\end{array}
\end{equation}
For a given set of momentum space parameters in $H_{3D}$, we can obtain corresponding experimental parameters to emulate the same system evolution over time $t_\text{eff}$.

In the quantum simulation experiment the initial state of the system is an eigenstate of $\gamma_{0} = \sigma_z \tau_z$, $\left|00 \right \rangle$, which is obtained by applying a green laser pulse. It corresponds to a deep quench along $\gamma_0$. Deep quenches along other axes are realized by either applying a microwave or radio-frequency pulse to prepare the system onto the eigenstates of $\gamma_{1,2,3}$, as illustrated in Fig. \ref{fig:experiment} (b).
For post-quench evolution, we plot the corresponding experimental circuit in Fig. \ref{fig:quantum circuit} (a). There are three steps in the post-quench operations:
(i) Rotate the nuclear spin along $-y$ axis for an angle $\theta$.
(ii) Apply a microwave pulse with a driving strength of $\omega_0$ and a phase of $\varphi$ during the evolution time of $t_\text{eff}$.
(iii) Rotate back the nuclear spin along $y$ axis for the same angle $\theta$. The net effect of this whole process is identical to the evolution of the system under $H_{3D}$ during an evolution time of $t$.

\subsection{Time-averaged spin polarization}
In the experiment all spin polarizations are finally measured in $z$ basis of electron and nuclear spins, where a population measurement $p_{\left|i,j \right \rangle}$ with $i, j = 0, 1$ can be obtained through the optical readout.
For the case of $\gamma_{0} = \sigma_z \tau_z$, which is already in the $z$ basis, the corresponding spin polarization is essentially $p_{\left|00 \right \rangle} - p_{\left|01 \right \rangle} - p_{\left|10 \right \rangle} + p_{\left|11 \right \rangle}$.
When we measure the spin polarization $\gamma_{1(2)} = \sigma_{x(y)} \otimes \mathds{1}$, we apply a $\pi / 2$ rotation on the electron spin about $-y$ and $x$ axes to map the $\sigma_{x}$ and $\sigma_{y}$ components to $\sigma_z$, respectively.
The corresponding spin polarization is given by $p_{\left|00 \right \rangle} + p_{\left|01 \right \rangle} - p_{\left|10 \right \rangle} - p_{\left|11 \right \rangle}$. Similarly, for the $\gamma_3 = \sigma_z \tau_x$ readout, a $\pi / 2$ rotation on the nuclear spin about $-y$ axis transforms the $\gamma_3$ readout to the $\gamma_0$ readout. These operations are depicted in Fig. \ref{fig:experiment} (b).

The population readout needs to record the photoluminescence (PL) photon count of the spin state, which is the average of all four levels weighted by their populations ($n_{tot} = n_1 p_{|00>} + n_2 p_{|01>} + n_3 p_{|10>} + n_4 p_{|11>}$). In order to obtain the populations, we apply different RF and MW pulses to produce equations of their different linear combinations and solve them. In Fig. \ref{fig:quantum circuit} (b) and (c) we show the sequences of the population measurement, and the corresponding equations are given as
\begin{equation}\label{}
\left( {\begin{array}{*{20}{c}}
{{n_1}}&{{n_2}}&{{n_3}}&{{n_4}}\\
{{n_3}}&{{n_4}}&{{n_1}}&{{n_2}}\\
{{n_2}}&{{n_1}}&{{n_3}}&{{n_4}}\\
{{n_3}}&{{n_4}}&{{n_2}}&{{n_1}}
\end{array}} \right)\left( {\begin{array}{*{20}{c}}
{{p_{\left| {00} \right\rangle }}}\\
{{p_{\left| {01} \right\rangle }}}\\
{{p_{\left| {10} \right\rangle }}}\\
{{p_{\left| {11} \right\rangle }}}
\end{array}} \right) = \left( {\begin{array}{*{20}{c}}
{{n_0}}\\
{{n_{MW}}}\\
{{n_{RF}}}\\
{{n_{MW + RF}}}
\end{array}} \right).
\end{equation}

We measure and average the spin polarization $\left\langle {{\gamma _i}\left( k \right)} \right\rangle$ under quenching along the $j$ direction over a series of time to obtained the time-averaged spin polarization ${\overline {\left\langle {{\gamma _i}\left( k \right)} \right\rangle } _j}$. For different polarizations and quench directions, we choose the same simulation time to keep consistency in experiments. The time range is chosen from $0$ to the maximum
\begin{equation}\label{}
{t_{\max }} = \frac{2}{{\sqrt 3 {t_{so}}\sin \left[ {\arccos \left( {3{m_0}/{t_0}} \right)} \right]}}.
\end{equation}

The shot noise of the optical readout leads to the dominant error in our quantum simulation, and yields a normal distribution of the photon counts with a mean of $N$ and a standard deviation of $\sqrt{N}$. We set $N$ between $2000$ and $3000$ for a fixed $20000$ repetitions of each sequence.

In Fig. \ref{fig:experimental data} we show a typical experimental result, where the parameters are set as $m_0 = 1.4t_0$, $t_{so} = 0.2t_0$, and the momentum $(k_x, k_y, k_z) = (0, 0, -\pi + \arccos 0.6)$ on the $3$-BIS. We perform quenches along the $j = 1$ and $2$ directions, and measure the spin polarizations $\gamma_i$ with $i = 1$, $2$, and $3$. Their time-averaged results are $\overline {\langle \gamma_1 \rangle}_1 = -0.0263 \pm 0.0170$, $\overline {\langle \gamma_2 \rangle}_2 = -0.0401 \pm 0.0132$, and $\overline {\langle \gamma_3 \rangle}_2 = 0.0165 \pm 0.0426$, which are very close to their theoretical values of zero.


\begin{thebibliography}{99}

\bibitem{PRL45.494}
K. von Klitzing, G. Dorda, and M. Pepper,
\textit{New method for high-accuracy determination of the fine-structure constant based on quantized Hall resistance},
{Phys. Rev. Lett.} \textbf{45}, 494 (1980).

\bibitem{PRL48.1559}
D. C. Tsui, H. L. Stormer, and A. C. Gossard,
\textit{Two-dimensional magnetotransport in the extreme quantum limit},
{Phys. Rev. Lett.} 48, 1559 (1982).

\bibitem{bookTQHE}
R. Prange, S. Girvin,
\textit{The Quantum Hall Effect} (Springer Verlag, Berlin, Germany, 1990).

\bibitem{RMP82.3045}
M. Z. Hasan and C. L. Kane,
\textit{Colloquium: topological insulators},
{Rev. Mod. Phys.} \textbf{82}, 3045 (2010).

\bibitem{RMP83.1057}
X. L. Qi and S. C. Zhang,
\textit{Topological insulators and superconductors},
{Rev. Mod. Phys.} \textbf{83}, 1057 (2011).

\bibitem{Yan2012}
B. Yan and S.-C. Zhang,
\textit{Topological materials},
{Rep. Prog. Phys.} \textbf{75}, 096501 (2012).

\bibitem{Chiu2016}
C.-K. Chiu, J. C. Y. Teo, A. P. Schnyder, and S. Ryu,
\textit{Classification of topological quantum matter with symmetries},
{Rev. Mod. Phys.} \textbf{88}, 035005 (2016).

\bibitem{Yan2017}
B. Yan and C. Felser,
\textit{Topological Materials: Weyl Semimetals},
{Annu. Rev. Condens. Matter Phys.} \textbf{8}, 337 (2017).

\bibitem{NPB922.62}
L. Kong, X.-G. Wen, and H. Zheng,
\textit{Boundary-bulk relation in topological orders},
{Nucl. Phys. B} \textbf{922}, 62 (2017).

\bibitem{CMP345.675}
V. Mathai and G. C. Thiang,
\textit{T-Duality Simplifies Bulk-Boundary Correspondence},
{Commun. Math. Phys.} \textbf{345}, 675 (2016).

\bibitem{JGP124.421}
K. C. Hannabuss,
\textit{T-duality and the bulk-boundary correspondence},
{J. Geom. Phys.} \textbf{124}, 421 (2018).

\bibitem{Nat547.298}
B. Bradlyn \textit{et al.},
\textit{Topological quantum chemistry},
{Nature} \textbf{547}, 298 (2017).

\bibitem{PRB61.10267}
N. Read and D. Green,
\textit{Paired states of fermions in two dimensions with breaking of parity and time- reversal symmetries and the fractional quantum Hall effect},
{Phys. Rev. B} \textbf{61}, 10267 (2000).

\bibitem{PU44.131}
A. Y. Kitaev,
\textit{Unpaired Majorana fermions in quantum wires},
{Phys.-Usp.} \textbf{44}, 131 (2001).

\bibitem{RPP75.76501}
J. Alicea,
\textit{New directions in the pursuit of Majorana fermions in solid state systems},
{Rep. Prog. Phys.} \textbf{75}, 76501 (2012).

\bibitem{RMP87.137}
S. R. Elliott and M. Franz,
\textit{Colloquium: Majorana Fermions in nuclear, particle and solid-state physics},
{Rev. Mod. Phys.} \textbf{87}, 137 (2015).

\bibitem{RPP80.076501}
M. Sato and Y. Ando,
\textit{Topological superconductors: a review},
{Rep. Prog. Phys.} \textbf{80}, 076501 (2017).

\bibitem{Nat452.970}
D. Hsieh \textit{et al.},
\textit{A topological Dirac insulator in a quantum spin Hall phase},
{Nature} \textbf{452}, 970 (2008).

\bibitem{NP5.598}
Y. Xia \textit{et al.},
\textit{Observation of a large-gap topological-insulator class with a single Dirac cone on the surface},
{Nat. Phys.} \textbf{5}, 398 (2009).

\bibitem{Science340.167}
C. Z. Chang \textit{et al.},
\textit{Experimental observation of the quantum anomalous hall effect in a magnetic topological insulator},
{Science} \textbf{340}, 167 (2013).

\bibitem{Science318.766}
M. K\"{o}nig \textit{et al.},
\textit{Quantum spin hall insulator state in hgte quantum wells},
{Science} \textbf{318}, 766 (2007).

\bibitem{PRX5.031013}
B. Q. Lv \textit{et al.},
\textit{Experimental Discovery of Weyl Semimetal TaAs},
{Phys. Rev. X} \textbf{5}, 031013 (2015).

\bibitem{Science349.613}
S.-Y. Xu \textit{et al.},
\textit{Discovery of a Weyl fermion semimetal and topological Fermi arcs},
{Science} \textbf{349}, 613 (2015).

\bibitem{LiuX2019}
X.-H. Pan, K. J. Yang, L. Chen, G. Xu, C. X. Liu, and X. Liu,
\textit{Lattice-Symmetry-Assisted Second-Order Topological Superconductors and Majorana Patterns},
{Phys. Rev. Lett.} \textbf{123}, 156801 (2019).

\bibitem{Science357.61}
W. A. Benalcazar, B. A. Bernevig, and T. L. Hughes,
\textit{Quantized electric multipole insulators},
{Science} \textbf{357}, 61 (2017).

\bibitem{PRL119.246401}
J. Langbehn, Y. Peng, L. Trifunovic, F. von Oppen, and P. W. Brouwer,
\textit{Reflection-Symmetric Second-Order Topological Insulators and Superconductors},
{Phys. Rev. Lett.} \textbf{119}, 246401 (2017).

\bibitem{PRL119.246402}
Z. Song, Z. Fang, and C. Fang,
\textit{$(d-2)$-Dimensional Edge States of Rotation Symmetry Protected Topological States},
{Phys. Rev. Lett.} \textbf{119}, 246402 (2017).

\bibitem{PRB92.085126}
R.-J. Slager, L. Rademaker, J. Zaanen, and L. Balents,
\textit{Impurity-bound states and Green's function zeros as local signatures of topology},
{Phys. Rev. B} \textbf{92}, 085126 (2015).

\bibitem{SA4.eaat0346}
F. Schindler, \textit{et al.},
\textit{Higher-order topological insulators},
{Sci. Adv.} \textbf{4}, eaat0346 (2018).

\bibitem{PRB97.205136}
E. Khalaf,
\textit{Higher-order topological insulators and superconductors protected by inversion symmetry},
{Phys. Rev. B} \textbf{97}, 205136 (2018).

\bibitem{PRX9.011012}
L. Trifunovic and P. W. Brouwer,
\textit{Higher-Order Bulk-Boundary Correspondence for Topological Crystalline Phases},
{Phys. Rev. X} \textbf{9}, 011012 (2019).

\bibitem{SB63.1385}
L. Zhang, L. Zhang, S. Niu, and X.-J. Liu,
\textit{Dynamical classification of topological quantum phases},
{Science Bulletin} \textbf{63}, 1385 (2018).

\bibitem{PRA99.053606}
L. Zhang, L. Zhang, and X.-J. Liu,
\textit{Dynamical detection of topological charges},
{Phys. Rev. A} \textbf{99}, 053606 (2019).

\bibitem{Zhang2019b}
L. Zhang, L. Zhang, and X.-J. Liu,
\textit{Characterizing topological phases by quantum quenches: A general theory},
{Phys. Rev. A} \textbf{100}, 063624 (2019).

\bibitem{Zhang2020}
L. Zhang, L. Zhang, and X.-J. Liu,
\textit{Unified theory to characterize Floquet topological phases by quench dynamics},
{Phys. Rev. Lett.} \textbf{125}, 183001 (2020).

\bibitem{dynamic1}
L. D'Alessio and M. Rigol,
\textit{Dynamical preparation of Floquet Chern insulators},
{Nat. Commun.} \textbf{6}, 8336 (2015).


\bibitem{dynamic2}
M. D. Caio, N. R. Cooper, and M. J. Bhaseen,
\textit{Quantum Quenches in Chern Insulators},
{Phys. Rev. Lett.} \textbf{115}, 236403 (2015).

\bibitem{dynamic3}
Y. Hu, P. Zoller, and J. C. Budich,
\textit{Dynamical Buildup of a Quantized Hall Response from Nontopological States},
{Phys. Rev. Lett.} \textbf{117}, 126803 (2016).

\bibitem{dynamic4}
B. Song \textit{et al.},
\textit{Observation of symmetry-protected topological band with ultracold fermions},
{Sci. Adv.} \textbf{4}, eaao4748 (2018).

\bibitem{dynamic5}
N. Fl\"aschner \textit{et al.},
\textit{Observation of dynamical vortices after quenches in a system with topology},
{Nat. Phys.} \textbf{14}, 265 (2018).

\bibitem{PRB98.205417}
L. Zhou and J. Gong,
\textit{Non-Hermitian Floquet topological phases with arbitrarily many real-quasienergy edge states},
{Phys. Rev. B} \textbf{98}, 205417 (2018).

\bibitem{dynamic6}
M. Tarnowski \textit{et al.},
\textit{Measuring topology from dynamics by obtaining the Chern number from a linking number},
{Nat. Commun.} \textbf{10}, 1728 (2019).

\bibitem{McGinley2019}
M. McGinley and N. R. Cooper,
\textit{Classification of topological insulators and superconductors out of equilibrium},
{Phys. Rev. B} \textbf{99}, 075148 (2019).

\bibitem{PRL124.160402}
H. Hu and E. Zhao,
\textit{Topological Invariants for Quantum Quench Dynamics from Unitary Evolution},
{Phys. Rev. Lett.} \textbf{124}, 160402 (2020).

\bibitem{PRL121.250403}
W. Sun \textit{et al.},
\textit{Uncover Topology by Quantum Quench Dynamics},
{Phys. Rev. Lett.} \textbf{121}, 250403 (2018).

\bibitem{PRL123.190603}
C.-R. Yi \textit{et al.},
\textit{Observing Topological Charges and Dynamical Bulk-Surface Correspondence with Ultracold Atoms},
{Phys. Rev. Lett.} \textbf{123}, 190603 (2019).

\bibitem{NP15.911}
B. Song \textit{et al.},
\textit{Observation of nodal-line semimetal with ultracold fermions in an optical lattice},
{Nature Physics} \textbf{15}, 911 (2019).

\bibitem{ZWang2020}
Z.-Y. Wang \textit{et al.},
\textit{Realization of ideal Weyl semimetal band in ultracold quantum gas with 3D Spin-Orbit coupling},
{Science} \textbf{372}, 271 (2021).

\bibitem{PRA100.052328}
Y. Wang \textit{et al.},
\textit{Experimental observation of dynamical bulk-surface correspondence in momentum space for topological phases},
{Phys. Rev. A} \textbf{100}, 052328 (2019).

\bibitem{Ji2020}
W. Ji \textit{et al.},
\textit{Quantum Simulation for Three-Dimensional Chiral Topological Insulator},
{Phys. Rev. Lett.} \textbf{125}, 020504 (2020).

\bibitem{Xin2020}
T. Xin, Y. Li, Y. A. Fan, X. Zhu, Y. Zhang, X. Nie, J. Li, Q. Liu, and D. Lu,
\textit{Quantum Phases of Three-Dimensional Chiral Topological Insulators on a Spin Quantum Simulator},
{Phys. Rev. Lett.} \textbf{125}, 090502 (2020).

\bibitem{Niu2020}
J. Niu \textit{et al.},
\textit{Simulation of Higher-Order Topological Phases and Related Topological Phase Transitions in a Superconducting Qubit}.
Preprint at https://arxiv.org/abs/2001.03933v1 and to appear in Science Bulletin (2021).

\bibitem{PRB78.195125}
A. P. Schnyder, S. Ryu, A. Furusaki, and A. W. W. Ludwig,
\textit{Classification of topological insulators and superconductors in three spatial dimensions},
{Phys. Rev. B} \textbf{78}, 195125 (2008).

\bibitem{Kitaev2009}
A. Kitaev,
\textit{Periodic table for topological insulators and superconductors},
{AIP Conf. Proc.} \textbf{1134}, 22 (2009).

\bibitem{PRB88.125129}
T. Morimoto and A. Furusaki,
\textit{Topological classification with additional symmetries from Clifford algebras},
{Phys. Rev. B} \textbf{88}, 125129 (2013).

\bibitem{PRB88.075142}
C.-K. Chiu, H. Yao, and S. Ryu,
\textit{Classification of topological insulators and superconductors in the presence of reflection symmetry},
{Phys. Rev. B} \textbf{88}, 075142 (2013).

\bibitem{PRB22.2099}
W. P. Su, J. R. Schrieffer, and A. J. Heeger,
\textit{Soliton excitations in polyacetylene},
{Phys. Rev. B} \textbf{22}, 2099 (1980).

\bibitem{PRL61.2015}
F. D. M. Haldane,
\textit{Model for a Quantum Hall Effect without Landau Levels: Condensed-Matter Realization of the ``Parity Anomaly"},
{Phys. Rev. Lett.} \textbf{61}, 2015 (1988).

\bibitem{Science294.823}
S. C. Zhang and J. P. Hu,
\textit{A Four-Dimensional Generalization of the Quantum Hall Effect},
{Science} \textbf{294}, 823 (2001).

\bibitem{FD}
H. B. Nielsen and M. Ninomiya,
\textit{Absence of neutrinos on a lattice: (I). Proof by homotopy theory},
{Nucl. Phys. B} \textbf{185}, 20 (1981).

\bibitem{PKNAWSA11.788}
L. E. J. Brouwer,
\textit{On continuous one-to-one transformations of surfaces into themselves},
{Proc. Kon. Nederl. Akad. Wetensch. Ser. A} \textbf{11}, 788-798 (1909).

\bibitem{Wu2016} Z. Wu, L. Zhang, W. Sun, X.-T. Xu, B.-Z. Wang, S.-C. Ji, Y. Deng, S. Chen, X.-J. Liu, and J.-W. Pan, Science {\bf 354}, 83 (2016).
\bibitem{Yi2019} C.-R. Yi, L. Zhang, L. Zhang, R.-H. Jiao, X.-C. Cheng, Z.-Y. Wang, X.-T. Xu, W. Sun, X.-J. Liu, S. Chen, and J.-W. Pan, Phys. Rev. Lett. {\bf 123}, 190603 (2019).

\bibitem{note1}
For example, the matrices can be taken as $\gamma_{0}=\sigma_{z}\otimes\tau_{z}$, $\gamma_{1}=\sigma_{x}\otimes\mathds{1}$,
$\gamma_{2}=\sigma_{y}\otimes\mathds{1}$, and $\gamma_{3}=\sigma_{z}\otimes\tau_{x}$.

\bibitem{PR528.1}
M. W. Doherty \textit{et al.},
\textit{The nitrogen-vacancy colour centre in diamond},
{Phys. Rep.} \textbf{528}, 1 (2013).

\bibitem{PRL102.057403}
V. Jacques \textit{et al.},
\textit{Dynamic Polarization of Single Nuclear Spins by Optical Pumping of Nitrogen-Vacancy Color Centers in Diamond at Room Temperature},
{Phys. Rev. Lett.} \textbf{102}, 057403 (2009).

\bibitem{Ktheory}
M. Karoubi,
\textit{K-theory: An introduction} (Springer, Berlin, 2008), vol. 226.

\bibitem{PRB55.1142}
A. Altland and M. R. Zirnbauer,
\textit{Nonstandard symmetry classes in mesoscopic normal-superconducting hybrid structures},
{Phys. Rev. B} \textbf{55}, 1142 (1997).

\bibitem{PRB90.165114}
K. Shiozaki and M. Sato,
\textit{Topology of crystalline insulators and superconductors},
{Phys. Rev. B} \textbf{90}, 165114 (2014).

\end{thebibliography}
\end{document}